# Evidence for the Independence of Waged and Unwaged Income, Evidence for Boltzmann Distributions in Waged Income, and the Outlines of a Coherent Theory of Income Distribution.

G. Willis* & J. Mimkes

Physics Department, University of Paderborn, 33096 Paderborn, Germany

*mimkes@physik.uni-paderborn.de*

*South Elmsall, W. Yorkshire, U.K.

*gmw@intonet.co.uk*

**Abstract**

Two sets of high quality income data are analysed in detail, one set from the UK, one from the USA. It is firstly demonstrated that both a log-normal distribution and a Boltzmann distribution can give very accurate fits to both these data sets. The absence of a power tail in the US data set is then discussed. Taken in conjunction with detailed evidence from the UK and Japanese income data, a strong case is made for the mathematically separate treatment of waged and unwaged income. The authors present a case for preferring the use of the Boltzmann distribution over the log-normal function, this leads to a brief review of the work of a number of researchers, which shows that a coherent theory for the distribution of all income can be postulated.

## Introduction

In mainstream economics the model for income distribution has traditionally been the log-normal. This has been based largely on the reasonable fit of the log-normal to the income data that has been collected. The log-normal distribution as a fit to economic data is not supported by an extant theoretical framework. The log-normal distribution has not been derived as a consequence of market-clearing equilibria (Walrasian, Marshallian etc models) as might reasonably be expected given the widespread acceptance of market-clearing equilibrium as the basis of modern economics, and the prescription of market-clearing theory by many economists and politicians as a good method of remedying income inequality.

The use of the log-normal as a best data fit is problematic. It has been known since the work of Pareto (1) as long ago as 1897 that income distributions have high end tails that follow power law decays. These distributions have been observed across a wide variety of different economic systems, and are persistently seen in many varied societies. Unfortunately the log-normal does not have such a power law tail, and as such forms a poor model to fit high end income data.

## Econophysics and Income Distribution

In recent years, the study of income distributions has gone through a small renaissance with new interest in the field shown by econophysicists such a Souma, Bouchaud & Mezard, Solomon & Richmond, Dragulescu & Yakovenko, Mimkes and others.

Wataru Souma has carried out extensive and detailed research into large data sets, primarily from Japanese income data (2,3). He has demonstrated with great clarity that income distributions consistently have a power tail decay at high income levels. He also demonstrates that the log-normal distribution gives a good fit to Japanese income data at lower income levels. The distributions for some of the years analysed show a smooth transition from the low income, log-normal curve to the high income, power tail distribution. However in some years there is a clear discontinuity between the two distributions (this point will be returned to later).

Independently, Jean-Phillipe Bouchaud & Marc Mezard (4), and Sorin Solomon & Peter Richmond (5) have proposed stochastic approaches using complexity theory to study income distribution. These have been very successful in explaining the origins of the power tails observed in income distributions, the first plausible theoretical frameworks to be proposed in the history of the study of income distribution. These frameworks have been less convincing in explaining the apparent log-normal distribution found at lower incomes.

Juergen Mimkes et al (6) have proposed a relatively straightforward maximum entropy equilibrium approach to explain income distribution for waged income, giving a Boltzmann distribution. This approach gives no power tail, and so does not explain the distributions seen at higher incomes.

Adrian Dragulescu and Victor Yakovenko (7) have also shown the presence of a Pareto power law tail in analysis of both US and UK income tax data. They have further proposed a Boltzmann-Gibbs exponential mid-section to the same data sets, while the fit proposed is highly suggestive, the quality of the source data leaves the results as inconclusive.

Finally, Arnab Chatterjee, Bikas Chakrabarti & S. S. Manna (8), Arnab Das & Sudhakar Yarlagadda (9) and have all used agent based approaches to income distribution based on gas-like market models. When saving propensity is fixed in these models, the outcome is a Gibbs like distribution. When the agents have a random saving propensity in the models, the distributions have power tails.

Amongst all of these papers, the problem of source data is a recurring one, most available data is collected by tax administration departments and is rough and ready in its nature. Income is generally aggregated in large bands, usually of differing widths. The number of points produced from any data set is usually insufficient for firm conclusions to be drawn, particularly at lower levels of income.

In this paper two unusually detailed data sets are analysed in an attempt to add some clarity to the ongoing debate on income distribution.

**United Kingdom NES Income Data**

The UK National Statistics Office (http://www.statistics.gov.uk/) runs an annual survey of weekly income of employees in the UK. This is known as the 'New Earnings Survey' or NES.

The NES takes a 1% sample of all employees in Great Britain (i.e. the United Kingdom excluding Northern Ireland). This includes full and part time employees. Individuals are identified through national insurance numbers and surveys of major employers. This makes the survey sample large and of very high quality and provenance.

It should be noted that the survey only includes wage earners, it excludes the unemployed, the self-employed, those earning less than the lowest tax threshold, and those living on private income. (These are collectively referred to as 'Unwaged income' for the purposes of this paper).

The data sets analysed in this paper come from the years 1992 to 2002 inclusive and show weekly incomes before tax.

Charts A, B & C show typical examples of the data, from 1992, 1997 and 2002. All data points from the data set have been shown, running from the £101-£110 band to the £1191-£1200 band. (Data above and below these levels has been lumped together in the data sets, and has not been included.

The graphs show a distribution which is clearly similar in form to the log-normal. The peak of the graph moves to the right, reflecting increasing overall wealth, as the years progress. The height of the peak also drops to compensate for the increasing width of the distribution.

The authors used the 'Solver' tool in Excel to attempt to fit two different functions to the full data sets in each of the years from 1992 to 2002.

The two functions used were,

The log-normal function:

$$F(x) = A*(EXP(-1*((LN(x)-M)*((LN(x)-M)))/(2*S*S)))/((x)*S*(2.5066)) \qquad (1)$$

And the Boltzmann function:

$$F(x) = B*(x-G)*(EXP(-P*(x-G))) \qquad (2)$$

In both cases the functions were fitted directly to the un-normalised data. This necessitated the use of a third parameter 'G' in the Boltzmann function to allow for an offset in the data. It is believed this offset can reasonably be ascribed to the presence of welfare payments and / or minimum wages, which would provide an artificial zero for the data.

The un-normalised log-normal is a three parameter function.

Three Parameter Fit

The data was fitted by using Solver to adjust the three parameters until a minimum was reached for the sum of the squares of the differences between the function values for F(x) and the actual data point values.

Charts D, E & F show typical results of these attempted fits to the data from 1992, 1997 and 2002.

The first thing to be noted from the results is that analysis by eye shows that both functions give excellent fits to the data.

Analysis of the results show that for both functions approximately 70% of the deviations from the curves have a value of less than 100. If it is assumed that the value of 100 is a reasonable error value, then the reduced chi-squared fit for the two curves can be calculated. These values are shown in chart G.

The reduced chi-squared values are all between values of 1.0 and 3.0. With 113 degrees of freedom this gives very good probability of fit between the theoretical curves and the actual data. For economic data this quality of fit is excellent.

The absolute values of reduced chi squared values need to be read with caution, given that the actual 'error' in the data is not known and has only been estimated. What is clear is that there is little to choose between the two functions shown. Generally the log-normal gives a slightly better fit than the Boltzmann, but not in all cases.

Charts H, I & J show typical results of the same data fits from 1992, 1997 and 2002. This time the y-axis show the logarithm of the count and fit, rather than the raw count.

These bring out some features which are common to all the data sets.

Firstly, there is a section of the data between approximately £400 and £800 that is almost linear on the log graph. This is indicative of a very strong exponential region of the raw data.

Secondly, all the graphs show a clear discontinuity at approximately £800 - £900. This supports previous work by others, most notably that of Souma, that there is a discontinuity between low end and high end income data. The log graphs show clearly that both the log-normal and the Boltzmann distributions have difficulties in modeling this high end data.

Thirdly, both the log-normal and the Boltzmann curves have problems modeling the first two or three data points near the origin. This is to be expected, as it is known that some employees below the tax threshold have been excluded from the survey.

Despite the high quality of the fit of the two curves to the data sets, the authors were concerned that the data points at the high and low ends were distorting the fit to the mid range data.

A second set of attempted data fits were carried out on a reduced data set for each year. The first three data points were excluded, the top third of the data points were also excluded; these being the points from the discontinuity and above. So the resulting data set for each year ran from £131-£140 to £791-800.

Charts K, L & M show the fits to the raw data for the three example years of 1992, 1997 & 2002. Charts N, O & P show the logarithms of the same data.

The fits of the theoretical curves to the data are now striking. It is also very noticeable that the two different functions fit each other almost exactly.

Analysis of the results show that for both functions approximately 70% of the deviations from the curves have a value of less than 150. If it is assumed that the value of 150 is a reasonable error value, then the reduced chi-squared fit for the two curves can be calculated. These values are shown in chart Q below.

The reduced chi-squared values are all between values of 0.5 and 1.2. With 69 degrees of freedom this gives extremely high likelihood of fit between the theoretical curves and the actual data.

Again the absolute values of reduced chi squared values need to be read with caution. It is however clear that there is little to choose between the two functions shown. This time the Boltzmann gives a slightly better fit than log-normal, but the difference between the two is too small to be given significance.

Two Parameter Boltzmann Fit

For the Boltzmann function only, the authors investigated the fit of the function to the data further. The family of Boltzmann functions is well understood mathematically and these functions can be normalised straightforwardly.

In this case it can be demonstrated that the function:

$$F(x) = B*(x-G)*(EXP(-P*(x-G))) \tag{2}$$

Can be re-written as:

$$F(x) = 10*No*P*P*(x-G)*(EXP(-P*(x-G))) \tag{3}$$

That is to say:

$$B = 10*No*P^2 \tag{4}$$

Here $No$ is the total number of people sampled from the population, this number $No$ is not a parameter, it is a quantity that is easily added up from the raw data. (The extra factor of ten arises as the bandwidth of the data is in units of 10: £101 - £110 etc.) By substituting for B in (2), function (3) has become a two parameter function.

The data was re-analysed and new best fits were obtained, again using solver in Excel to fit the data. In each case $No$ was calculated from the raw input data. The resulting fits were then two parameter fits, as only P and G were adjusted to produce the resultant fit. The fitting was again done with only the reduced data set where it is believed the Boltzmann distribution holds sway. It should be noted that $No$ was summed from the whole data set.

Charts R, S & T show the fits to the raw data for the three example years of 1992, 1997 & 2002.

The fits are not quite as good as those obtained with a three parameter fit, however the fits of the curves to the data is still excellent.

One Parameter Boltzmann Fit

It can be further demonstrated that if the data is modeled by a Boltzmann distribution, then the value of P can be independently related to G and information contained in the raw data. That is to say it can

be shown that:

P = 2 / (*Ko*/*No*-G)  (5)

Here *Ko* is the total sum of the number of people in each population band multiplied by the average income of this income band. Again *Ko* can be calculated directly from the raw data, without any reference to the graph. In this case we know that the value of *Ko* is likely to be slightly inaccurate for a Boltzmann fit. We have already noted above that the data includes a logarithmic tail of high values. These values are above the Boltzmann curve, so the calculated values of *Ko* from the data will be higher than the value that would give an ideal Boltzmann fit.

Leaving aside this reservation, it is possible to substitute equation (5) into equation (3) to produce the rather cumbersome function:

F(*x*) = 10\**No*\*(2/((*Ko*/*No*)-G))\*(2/((*Ko*/*No*)-G))\*(*x*-G)\*(EXP(-(2/((*Ko*/Pop)-G))\*(*x*-G)))  (6)

Here G is left as the only parameter in the function. For each year both *No* and *Ko* can be calculated directly from the raw data.

Once again the data was analysed and new best fits were obtained, again using solver in Excel to fit the data. The fitting was again done with only the reduced data set where it is believed the Boltzmann distribution holds sway. However *No* and *Ko* were summed from the whole data set.

Charts U, V & W show the fits to the raw data for the three example years of 1992, 1997 & 2002.

As can be seen from the charts, the fits to the data are still reasonably good, though clearly not as good as the two and three parameter fits. Given that there are known sources of error in the data, in the form of the extra 'logarithmic income', the fits are still surprisingly good.

Defined Parameter – Generated Boltzmann Function

It is possible to take this one stage further.

As discussed above, *No* & *Ko* can be calculated directly from the raw data. The parameter G was introduces as it was believed that there was an offset present in the data set. It is also possible to make an estimate of the value of G from the raw data. This was done by taking the first few points in the data series, and calculating a simple linear regression through these points back to the intercept on the *x* axis.

This value of G was then substituted directly into equation (6) above to produce the results shown in charts X, Y & Z for the three example years of 1992, 1997 & 2002.

In this case Solver was used in Excel only to fit the data for the linear regression to calculate G. For the years 1997 and 2002 the first 12 points were used from the data sets, for 1992 the first 6 points only are used. Clearly this choice was a subjective one, guided by what the authors thought was a representative linear section. Readers are invited to inspect charts X, Y & Z for themselves to decide whether they think the choice was reasonable.

As can be seen from the charts, the fits to the data are still reasonably good, though clearly not as good as the one, two or three parameter fits. Here there are known sources of error in both the calculated value of *Ko* and in the estimated value of G. Despite the deficiencies of the estimates of the scaling parameters, the fits are still surprisingly good.

(In practice, the best way to estimate G & P (and so *Ko*) would be to derive these from the two parameter fit above. However this is not the point of the paper.)

To recap, these final, defined parameter, graphs have been achieved in the following manner:

Firstly it has been assumed that the data is distributed as a Boltzmann distribution that has an offset from zero. Then the scaling constants have been directly calculated from the raw data. The offset G has been estimated from the first few points of the raw data. *No*, *Ko* and G have effectively defined P. And finally the points for the curve have been generated from equation (6) the formula for the Boltzmann distribution. These points have been plotted directly on the graph. In effect the curves have been defined from first principles using the bulk data; they have not been 'fitted' directly to the data.

**United States Waged Income Data**

The second data set analysed is from a 1992 survey giving proportions of workers earning particular wages in manufacturing and service industries (10).

The ultimate source of the data is the US Department of Labor; Bureau of Statistics, and so the provenance is believed to be of the highest quality. Unfortunately, enquiries by the authors have failed to reveal the details of the data, such as sample size and collection methodology.

The data was collected to give a comparison of the relative quality of employment in the manufacturing and service sectors. As with the UK data, the data is for waged income. (Unfortunately the data was given in proportions of the population, so *No* and *Ko* could not be calculated for this data.)

Although the sample size for the data is not known, the smoothness of the curves produced suggest that the samples were large, and that the data is of high statistical quality.

Figures AA & BB show the two data sets, along with the log-normal and Boltzmann curves generated using Excel's 'Solver'.

As can be seen by eye, the fit of both the log-normal and Boltzmann functions to the actual data is very good indeed.

Without a meaningful estimate of the errors in the data source, the absolute values of chi-squared are not possible to determine. The relative values of the sum of the squares of the errors for the two functions indicate that the log-normal function gives a slightly better fit to the manufacturing data, while the Boltzmann function gives a slightly better fit to the services data.

Figures CC & DD show the two data sets plotted on logarithmic scales.

Like the UK data, there appears to be a clear linear section in the central portion of the data, indicating an exponential section in the raw data.

What is much more interesting is that, beyond this section, the data heads rapidly negative on the logarithmic scale. This means it is heading rapidly to zero on the raw data graph. With these two distributions there is no sign whatsoever of the 'power tail' that is normally found in income distributions.

This leads to the first main conclusion that can be drawn from these data sets.

**Independence of Waged and Unwaged Income**

As discussed previously, Souma has done extensive research into income data sets, primarily from Japan. He has demonstrated clearly that high end income follows a power law, while the lower end appears to follow a log-normal distribution. More interestingly, while some years in the Japanese data show a continuous smooth distribution from high to low incomes, many years show a very clear discontinuity between the middle income section and the power tail.

The UK NES data in this paper follows the same pattern, with a high end tail that is discontinuous with the main distribution. It has also been shown that the low and medium incomes can be modeled equally well by either a log-normal or a Boltzmann function. This fits well with the studies above.

The US industry data in this paper is more tantalising. For some reason of the data selection used in the survey, it appears that the surveyors have effectively removed the contribution from the power tail, and have only measured the low to middle end of the distribution. The resultant distributions can be modeled completely with either the log-normal or the Boltzmann alone.

It is the belief of the authors that the discontinuities seen in the Japanese and UK data, together with the absence of a power tail in the US industry data, suggest that two separate mathematical regions are in operation. It is believed that rather than this being a split which blends from low to high income levels, it is actually a split that results from two economically different systems. That is that most people, those who are in contracted, waged, employment; have income levels which are determined by a mathematical function that is represented by either the log-normal or the Boltzmann.

A much smaller group of people, whose income is primarily derived from investment income, have their incomes dictated according to a power law. The cross over between the two groups is believed to be quite small, and it appears that 'unwaged' income comprises an 'add-on' 'tail' to the 'waged' income 'body'. It should be noted that although this tail will form a small proportion of the people, it will form a substantial part of the total income.

It appears that different economic rules apply for different types of income. The job market is relatively stable and is characterised by long term contracts and requires one form of mathematics. Stock and property markets are more volatile and need a more complex approach.

This is a subtle, but importantly different way of looking at income distribution. It has been clear for some time that two different mathematical functions are needed to explain the overall distribution. However it has implicitly been assumed that one function models low and median incomes, a second distribution governs higher incomes, with a transition zone between the two areas. The conclusion of this paper is that there are two substantially separate different distributions in existence, which overlap, and are additive when overall income is examined.

This view is supported by the modeling work of Chatterjee, Chakrabarti & Manna which gives different models of income according to the whether the propensity to save is fixed or not.

Subjectively this seems quite reasonable. For most people, even quite wealthy people, saving is largely restricted to investments in those people's own homes and pensions. In these cases the propensity to save is effectively fixed. Only a small number of people are sufficiently wealthy to be able to make choices over whether they wish to save or not.

**Modeling Waged Income**

With regard to how the low and medium incomes should be modeled, the data studied in this table presents an embarrassment of riches.

In the field of economics it is unusual to find a detailed data set that exactly matches a theoretical curve.

In this paper it has been demonstrated that both the log-normal and the Boltzmann distribution can give near perfect fits to data sets from two different countries.

The strong argument in favour of the Boltzmann is that it is fundamentally a single parameter curve and that its shape is essentially fixed. At low values it rises proportional to $x$, at high values it declines as an exponential. In the middle you get a peak where the two functions cancel each other out. What is important to note is that the ratio of the slopes on either side of the peak is effectively fixed.

This is not true of the log-normal, which varies in shape as sigma (S in eq (1) above) is varied.

For example figure EE gives an example of a log-normal that mimics the Boltzmann function on the low income side, but falls much more rapidly on the high income side, giving a more 'symmetrically' shaped peak.

In both the UK and US data studied above, it is highly coincidental that the log-normal that gives a perfect fit is the log-normal that perfectly mimics the Boltzmann distribution.

With the UK data it has been further demonstrated that the data can be very closely modeled by deriving the scaling factors from the raw data and using a two or one parameter fit. In extremis it has even been possible to model the UK data directly from analysis of the underlying data.

In scientific study it is normal to accept the simplest theory that fits the evidence, in this case a two parameter fit should be preferred to a three parameter fit.

If, in the future, a single data set is discovered that follows a log-normal that does not coincide with a Boltzmann, then the log-normal should be the preferred function. However, the Boltzmann is the simpler function, and two independent data sets have been found to follow this function. Until a data set is discovered that does not follow the Boltzmann function, it is more logical to use the Boltzmann function.

The authors also favour the Boltzmann for more partisan reasons. The authors believe that there is a straightforward and robust mathematical theory available to generate this distribution. The authors have assumed that the employment market is in a classical equilibrium (that is a physicist's classical thermodynamic equilibrium, not an economist's classical equilibrium) and that overall conservation principles can be used. This allows a simple maximum entropy statistical approach. This approach is fully discussed in a previous paper by the authors (6).

This equilibrium assumption would appear to have some merit. The UK income data covers a period of significant economic change, including the expulsion of the UK from the European Monetary System, the introduction of the minimum wage and the end of the dot.com boom, as well as varying rates of economic growth. Despite this the successive yearly distributions are very similar in shape.

Similarly the two curves for US industry in 1992 show almost identical distributions for manufacturing and services, despite the fundamentally different natures of these industries.

Despite enquiry with economists who are expert in the field of income distribution, the authors have not been able to find a theoretical basis for the use of the log-normal distribution in income distribution.

Despite the arguments above, it would be unwise to make a categoric assertion in favour of the Boltzmann distribution. Two sets of data is only just enough to support a theory. In this case verification is necessary, and in theory verification should be easy, though possibly expensive, to do. Similar statistical studies of waged income need to be done in other countries, and analysed to see which distribution is appropriate.

**A Coherent Theory of Income Distribution**

Notwithstanding the reservations in the previous section, it would appear that the combined work of a number of researchers has reached a point where econophysics can supply a complete and coherent explanation of income distribution supported by both theory and data.

The data analysis of Souma, Dragulescu & Yakovenko, and others has shown a very strong case for there being a power law tail at the high end of income data from Japan and the US.

For the low income end, this paper has analysed data to demonstrate that the Boltzmann distribution provides a very good fit to data from the UK and the US.

Analysis of all the above data taken together gives a good argument for the logical separation of income into two mathematically distinct areas.

For the area of waged income where long term contracts provide stability, and so equilibrium, a theoretical framework has been provided by Mimkes, Fruend & Willis.

The modeling of the power law tail, will however need non-equilibrium mathematics of the form already provided by Richmond, Solomon, Bouchaud & Mezard.

Finally the modeling work already carried out by Chatterjee, Chakrabarti, Manna, Das & Yarlagadda suggests that the above theoretical framework is a sound one.

**Conclusions**

In brief the following are the conclusions of this paper.

Firstly, as has been known for some time, the log-normal distribution is not a suitable model for overall income distribution.

Waged income and unwaged income appear to be effectively independent. In practice the two separate distributions produced by these mathematically different functions are juxtaposed by data collection techniques.

While the log-normal may provide a good model for waged income distribution, the Boltzmann distribution appears to provide simpler and more robust model than the log-normal. The Boltzmann model also has a theoretical grounding.

Further theoretical explanation, almost certainly drawn from complexity theory, is needed to explain the distribution of unwaged income.

By far the most important conclusion is that this area is ripe for further investigation. Taken together, the work of the various authors listed above has come close to giving a well based set of theories that give a full explanation of how income is distributed in human societies. More data collection is essential, with representative samples of waged and unwaged income being gathered across a range of different countries.

A

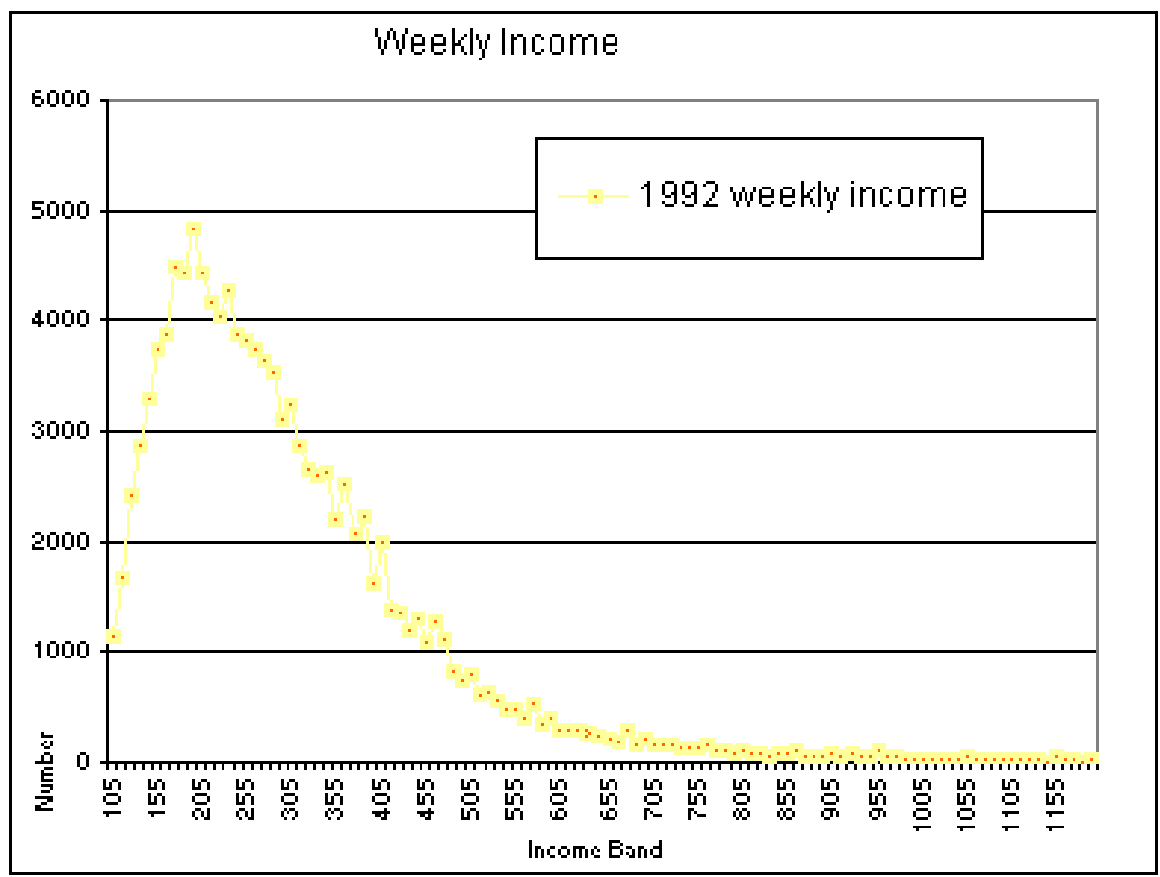

B

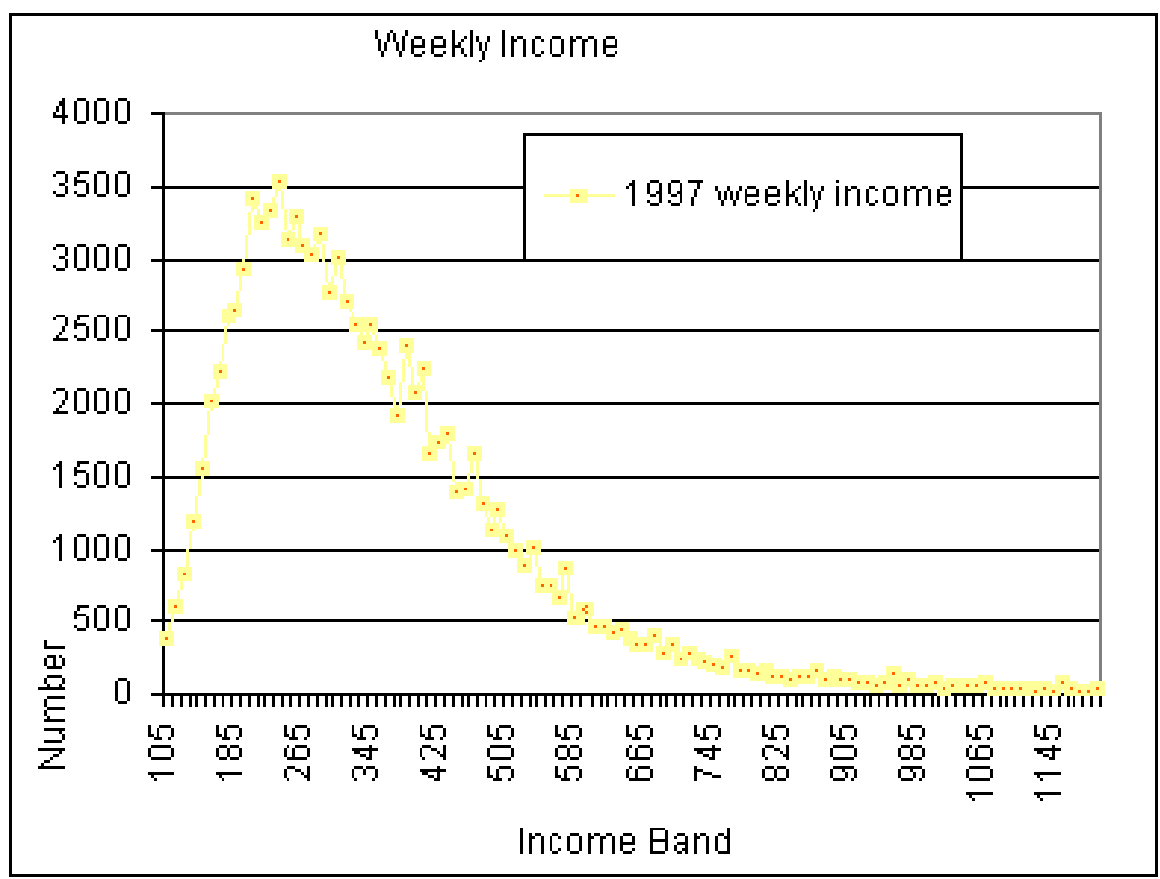

C

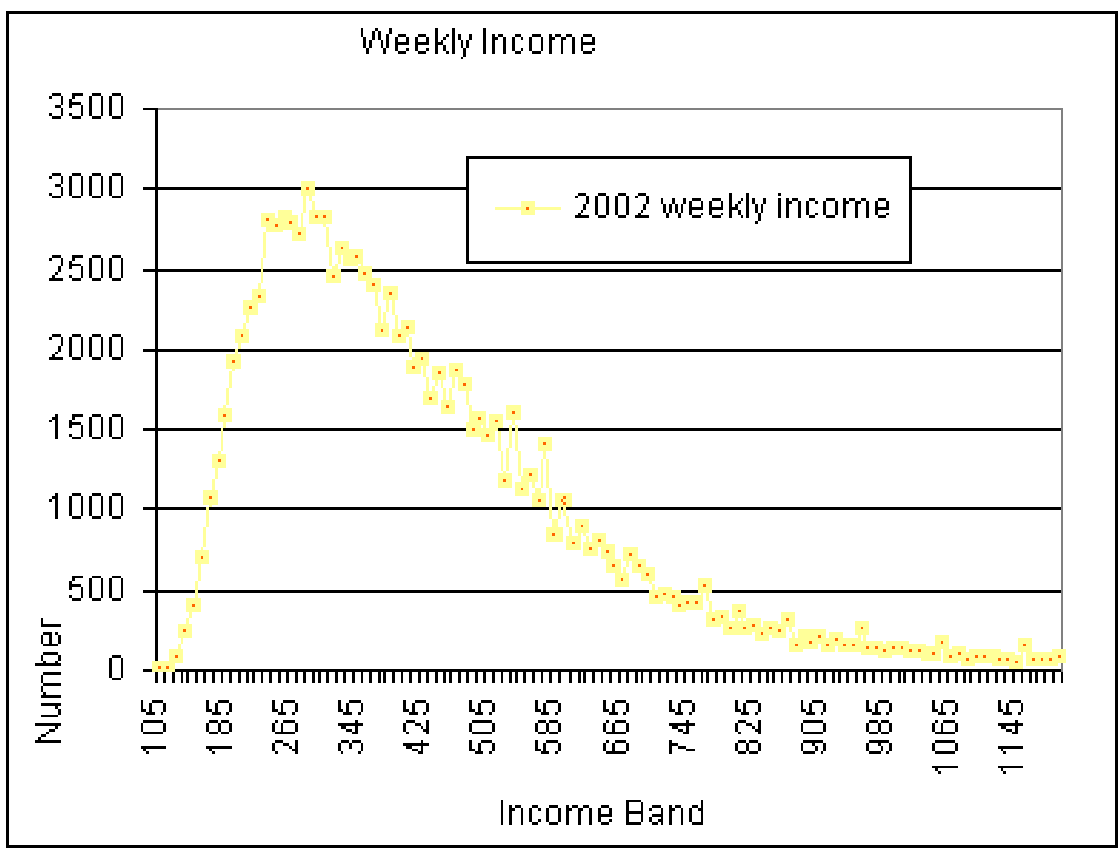

D

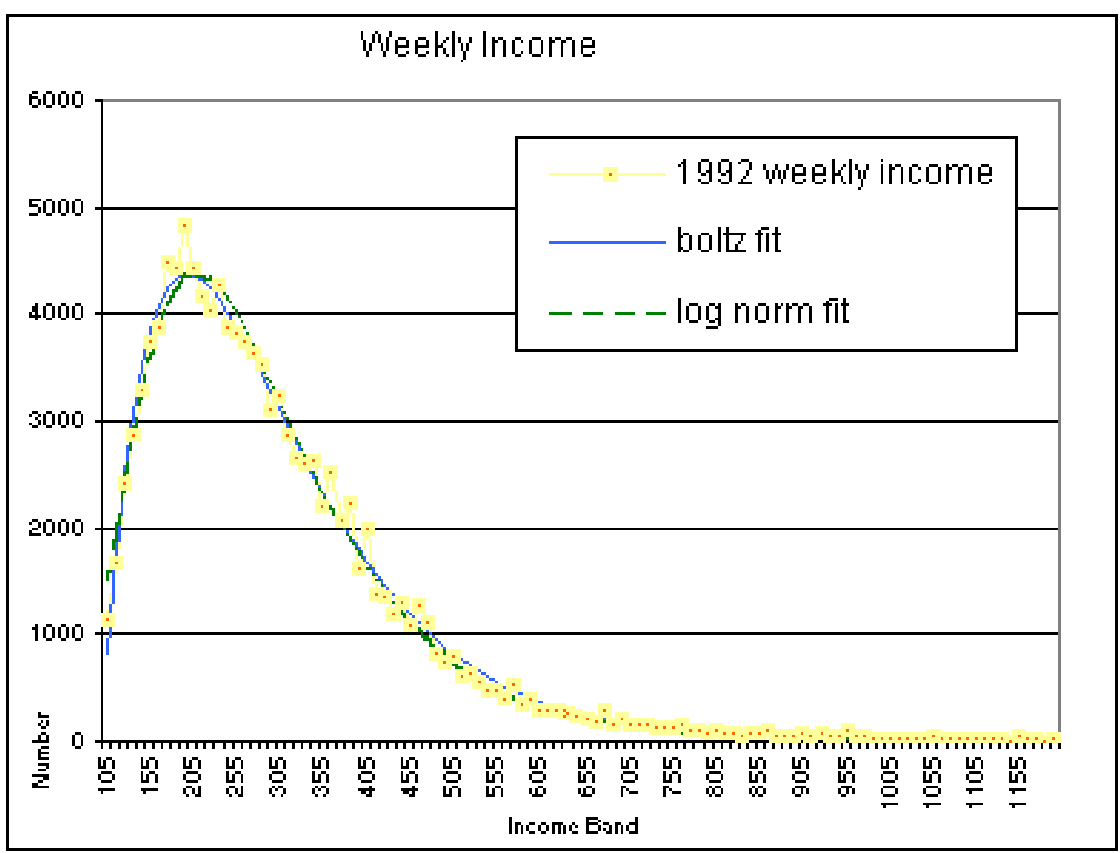

E

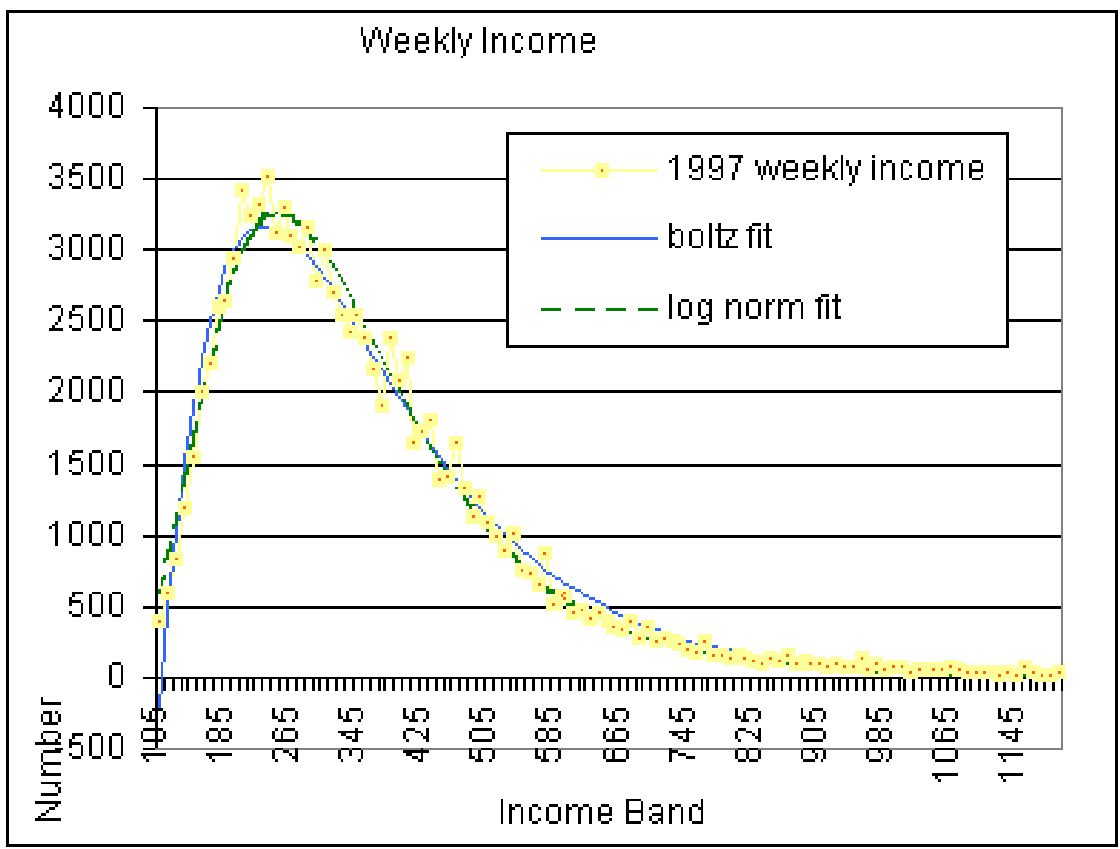

F

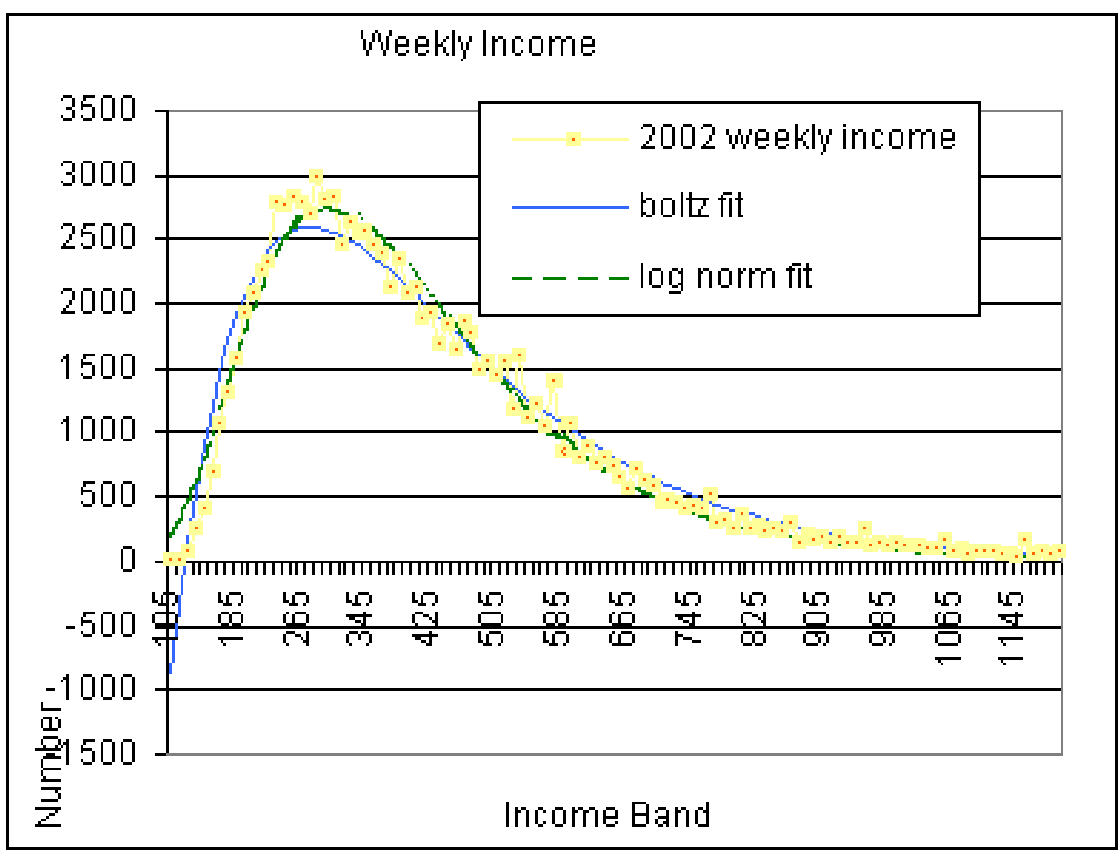

G

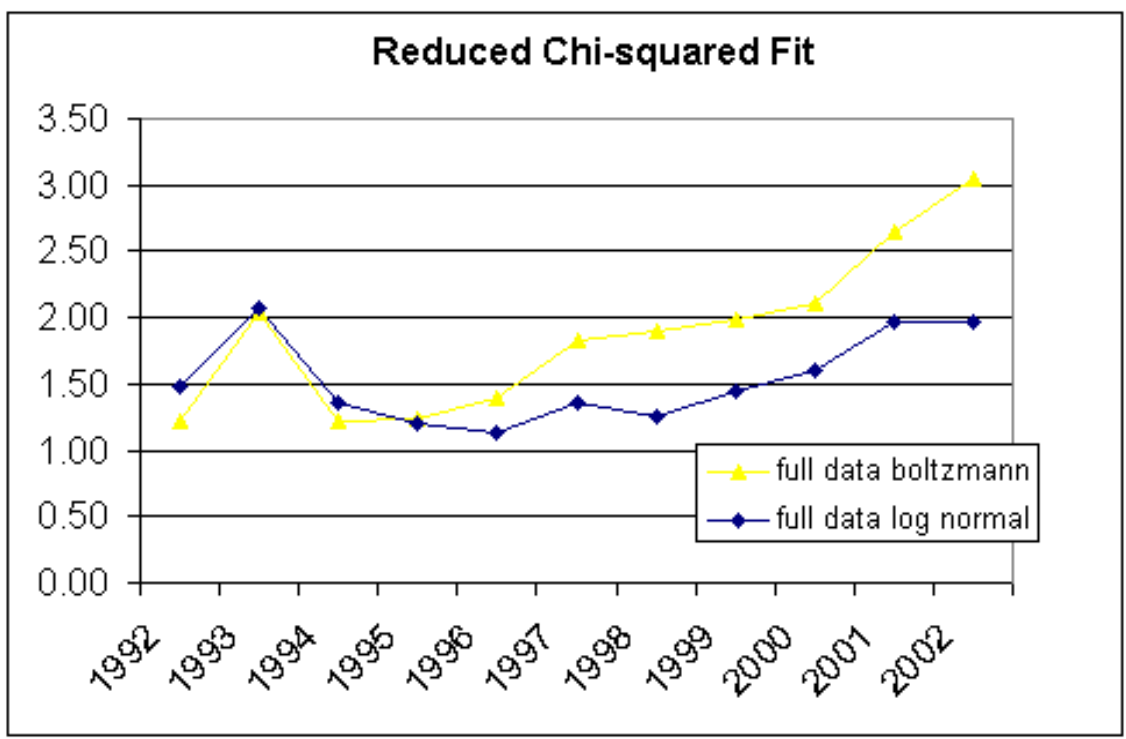

H

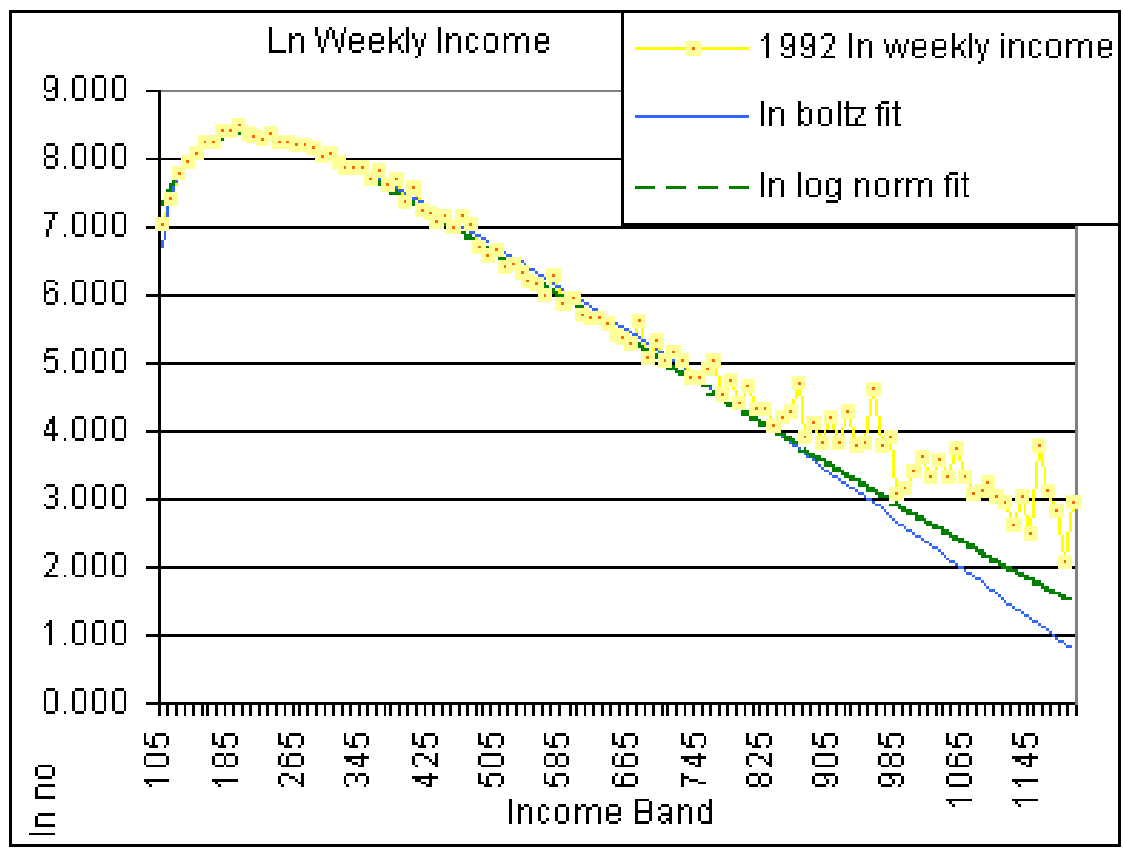

I

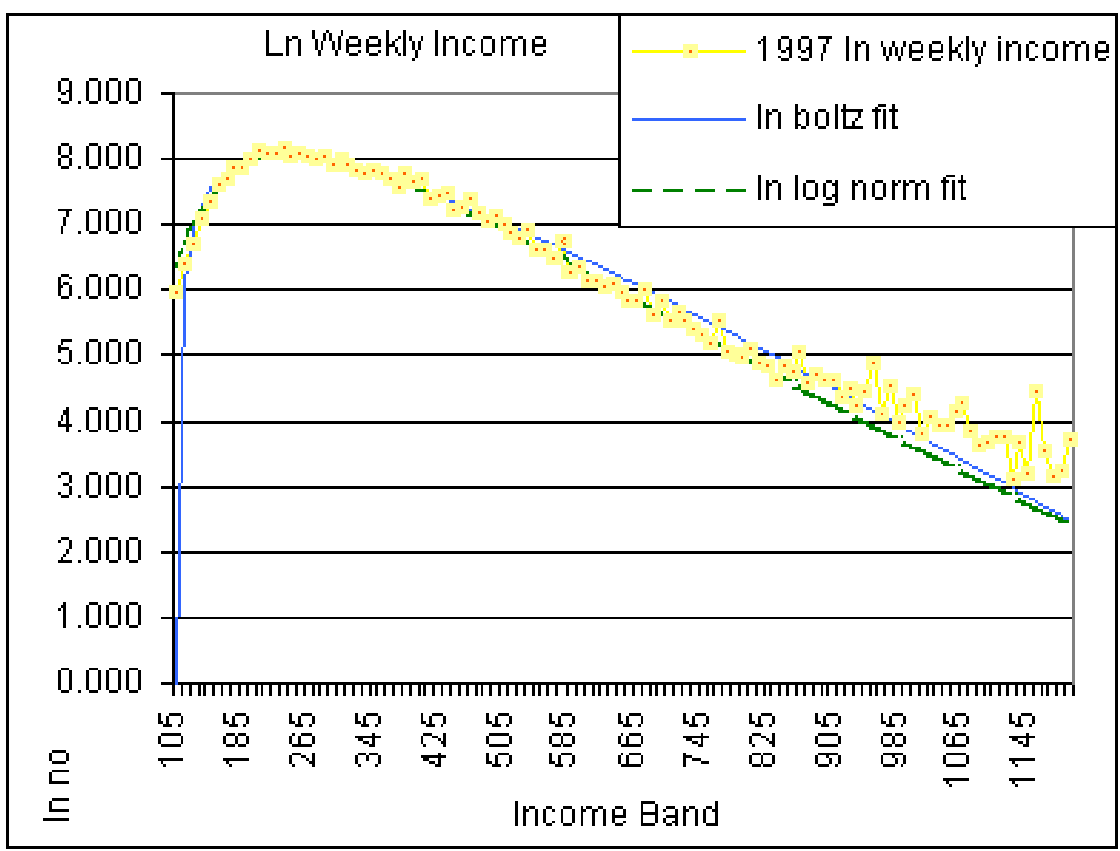

J

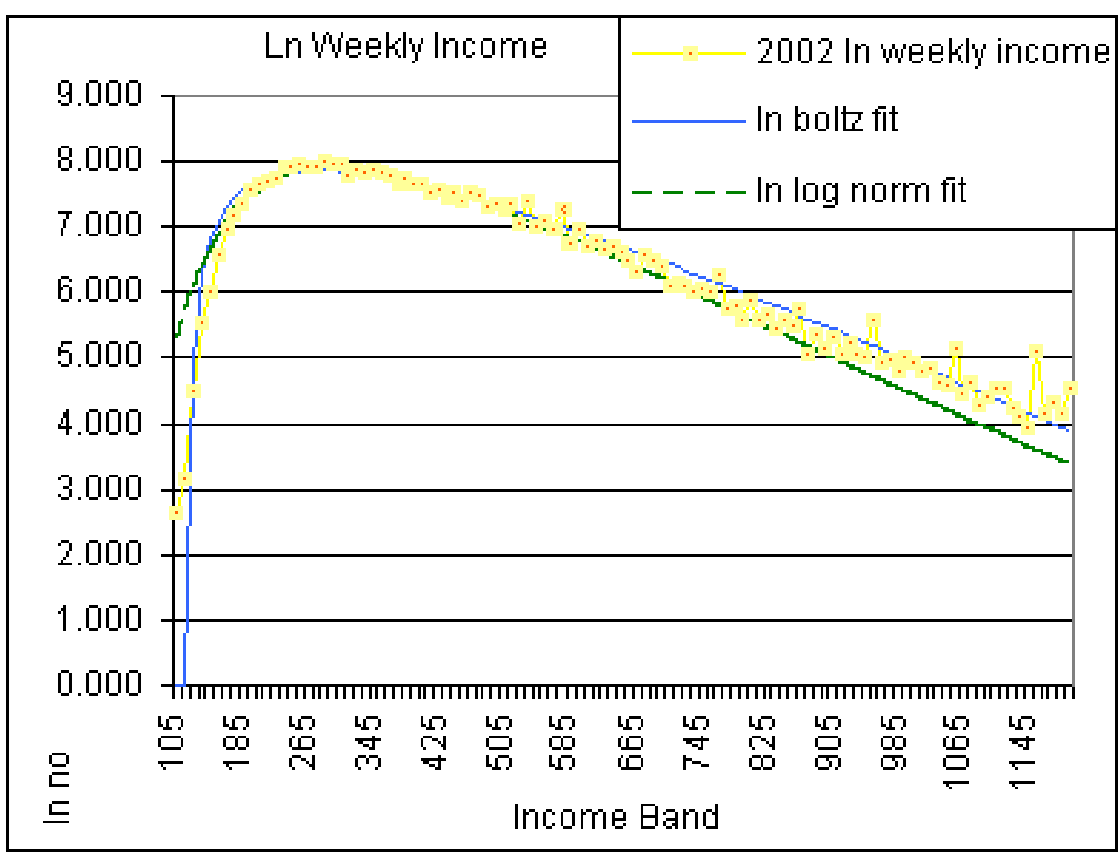

K

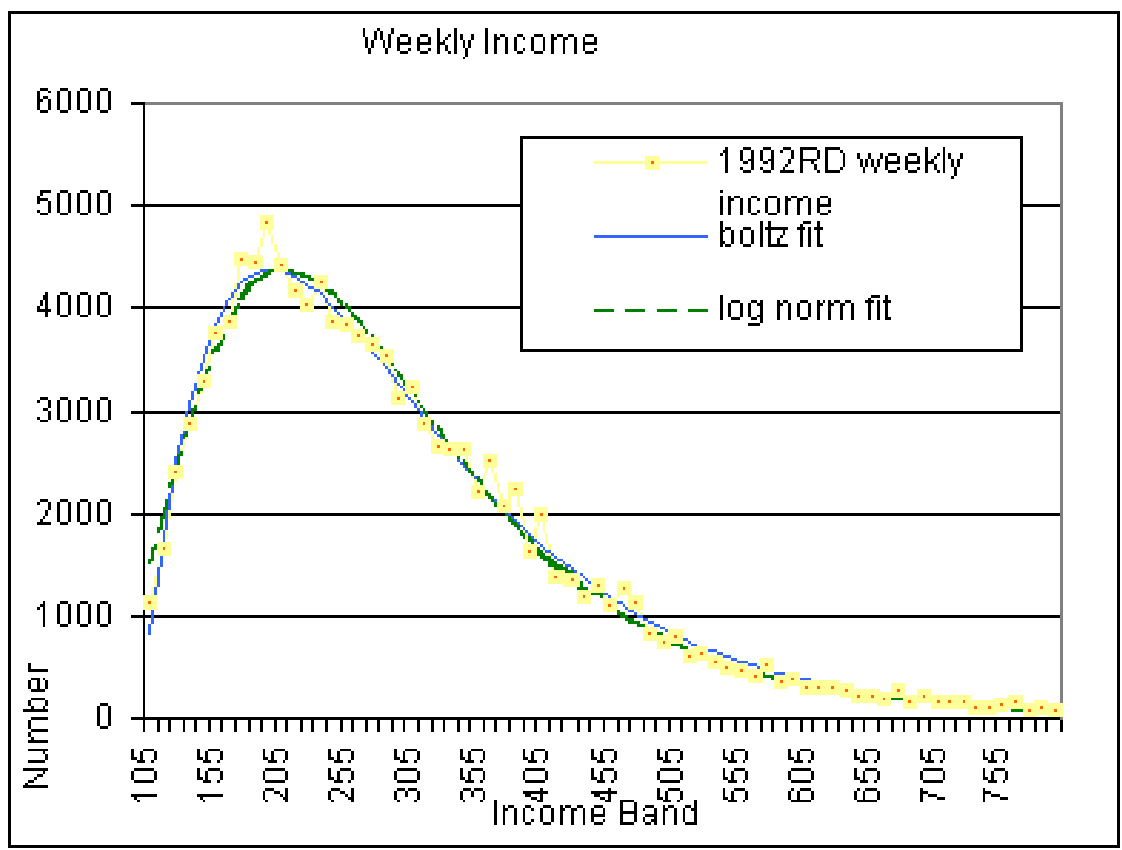

L

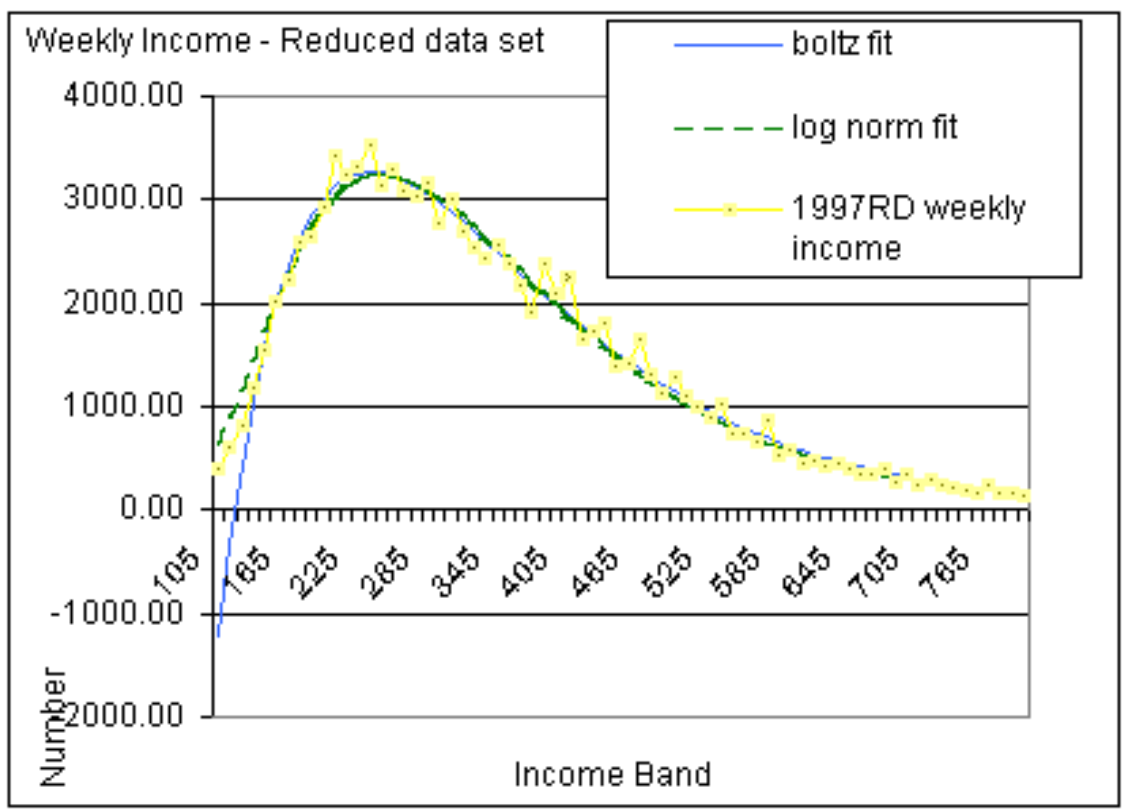

M

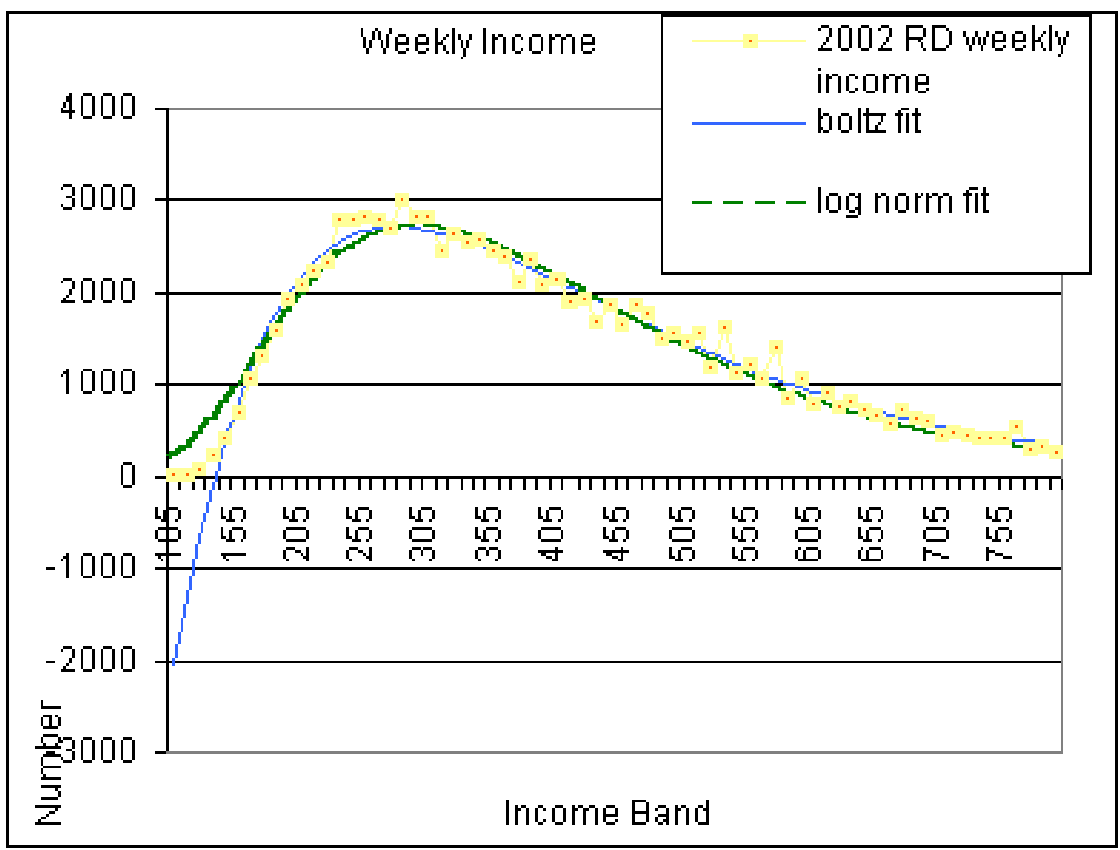

N

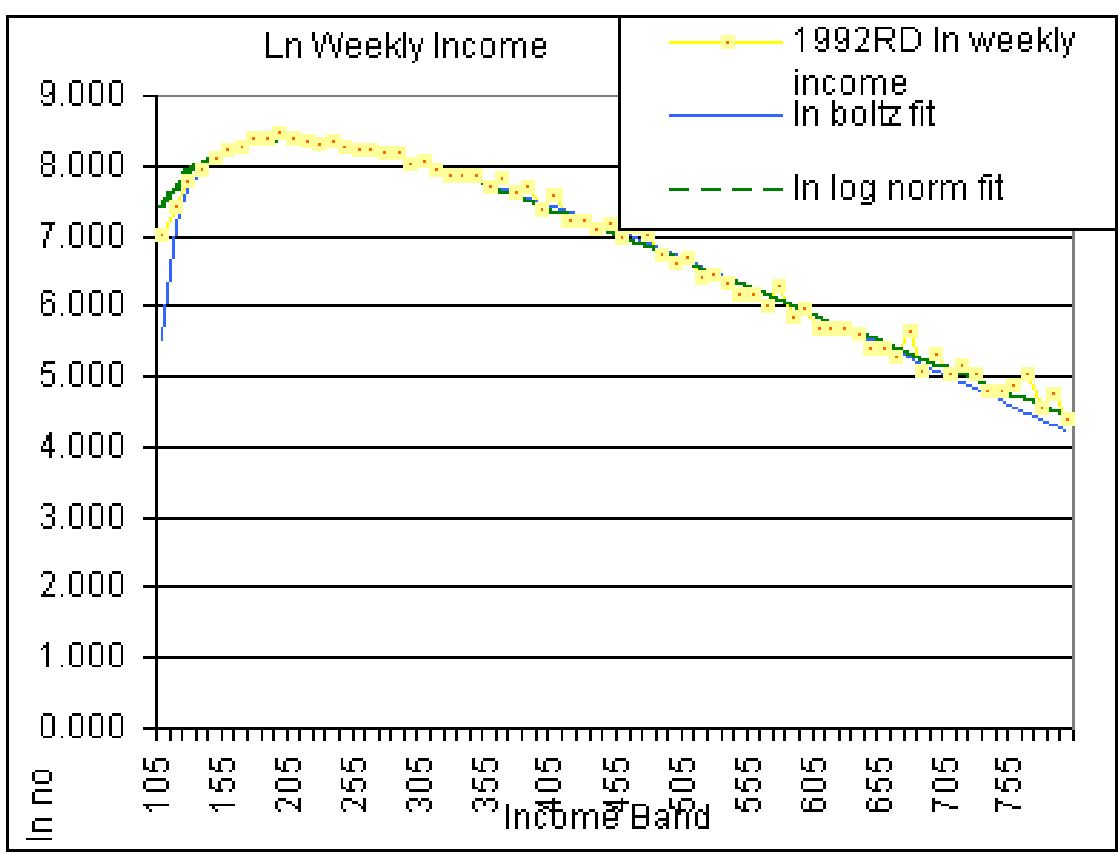

O

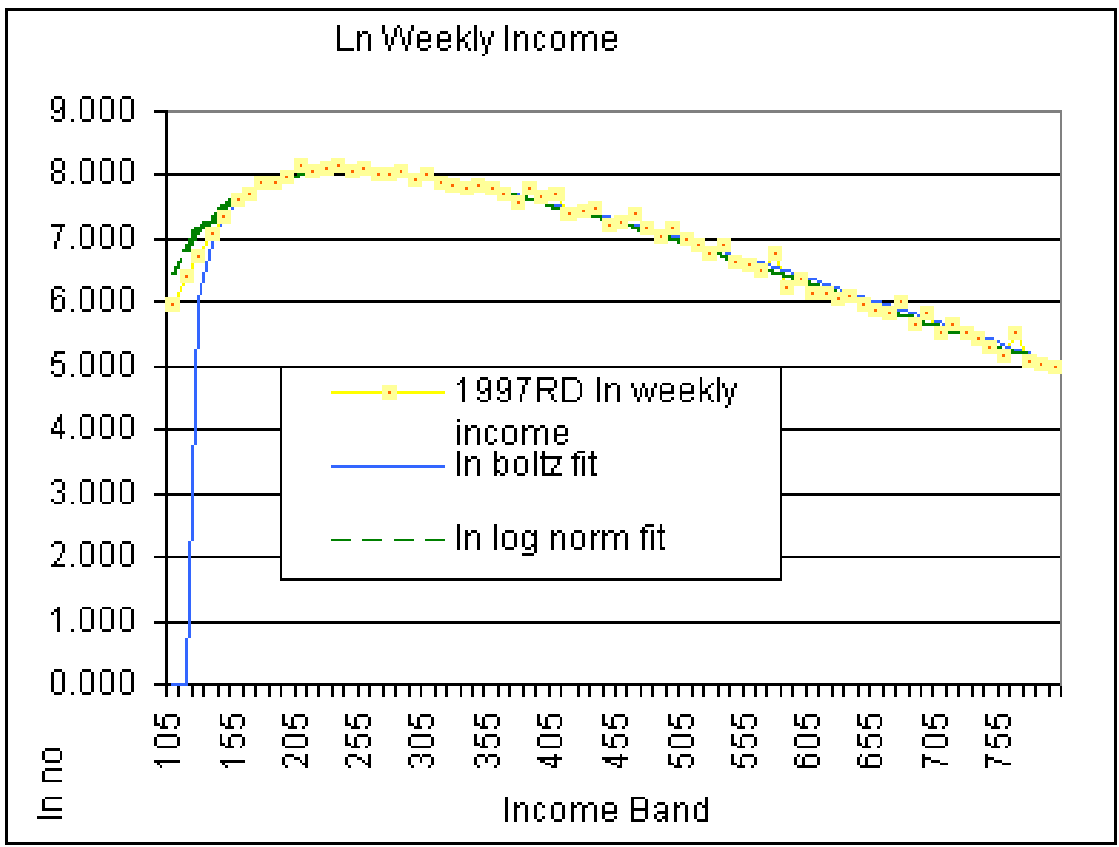

P

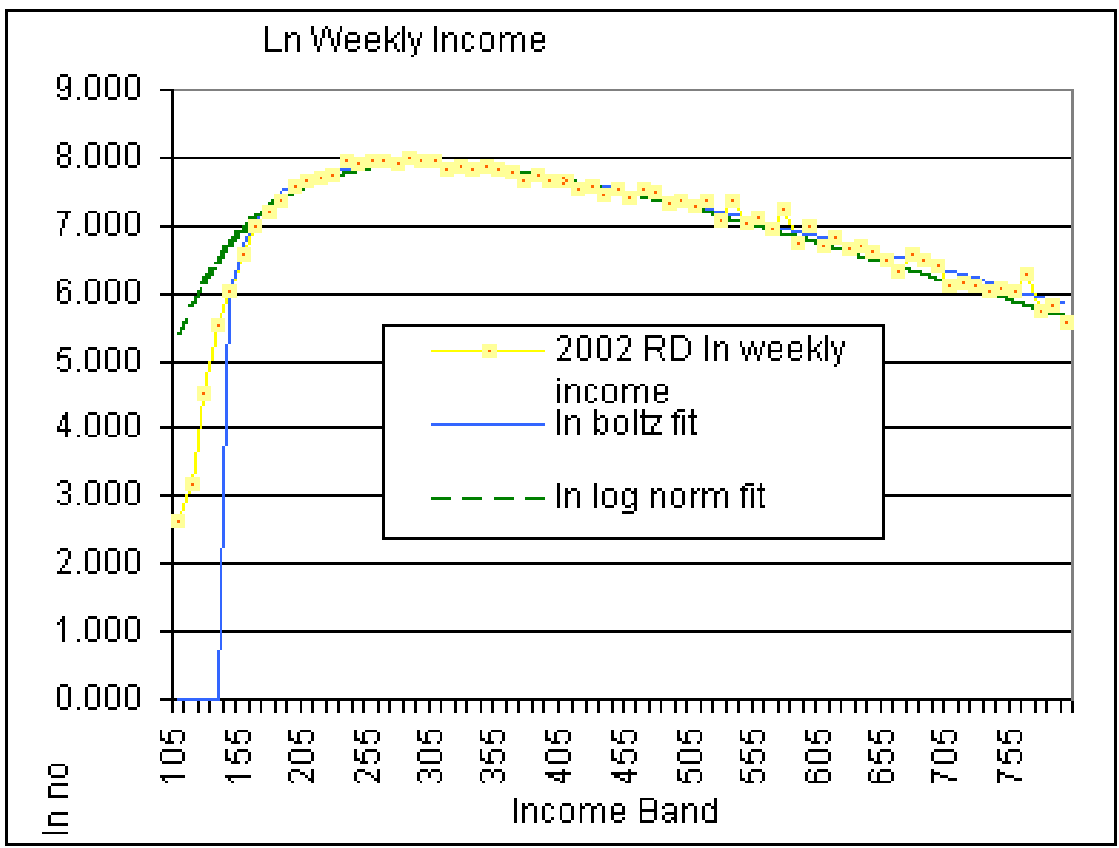

Q

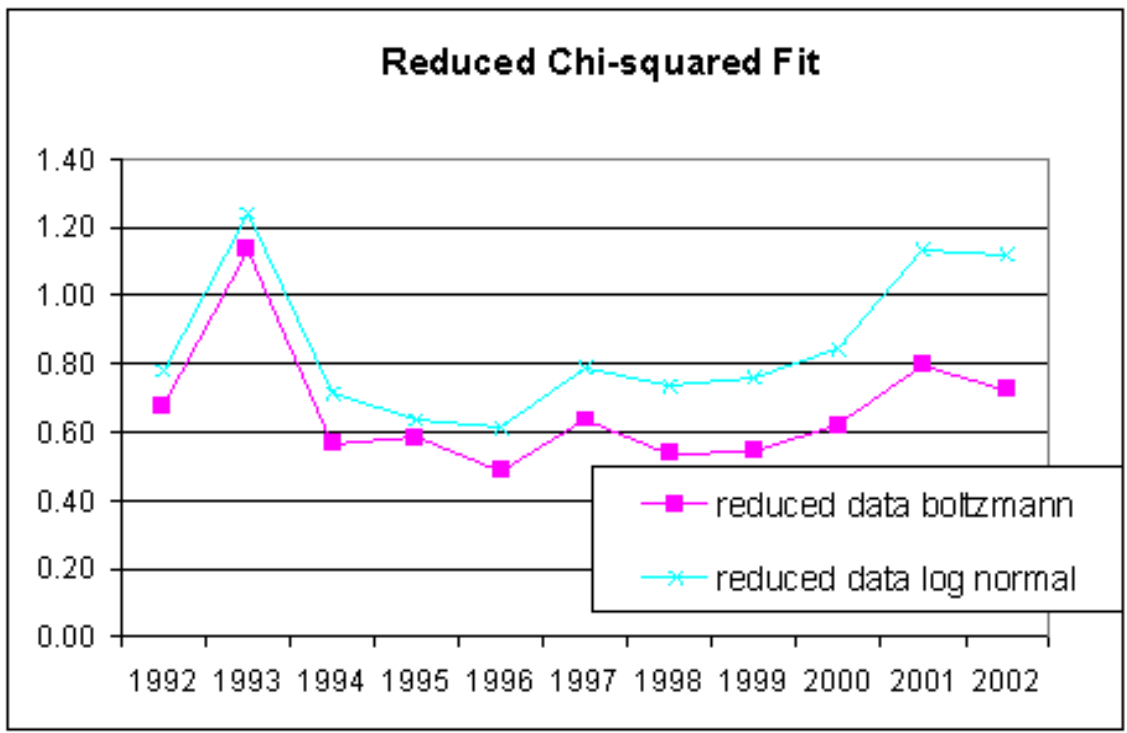

R

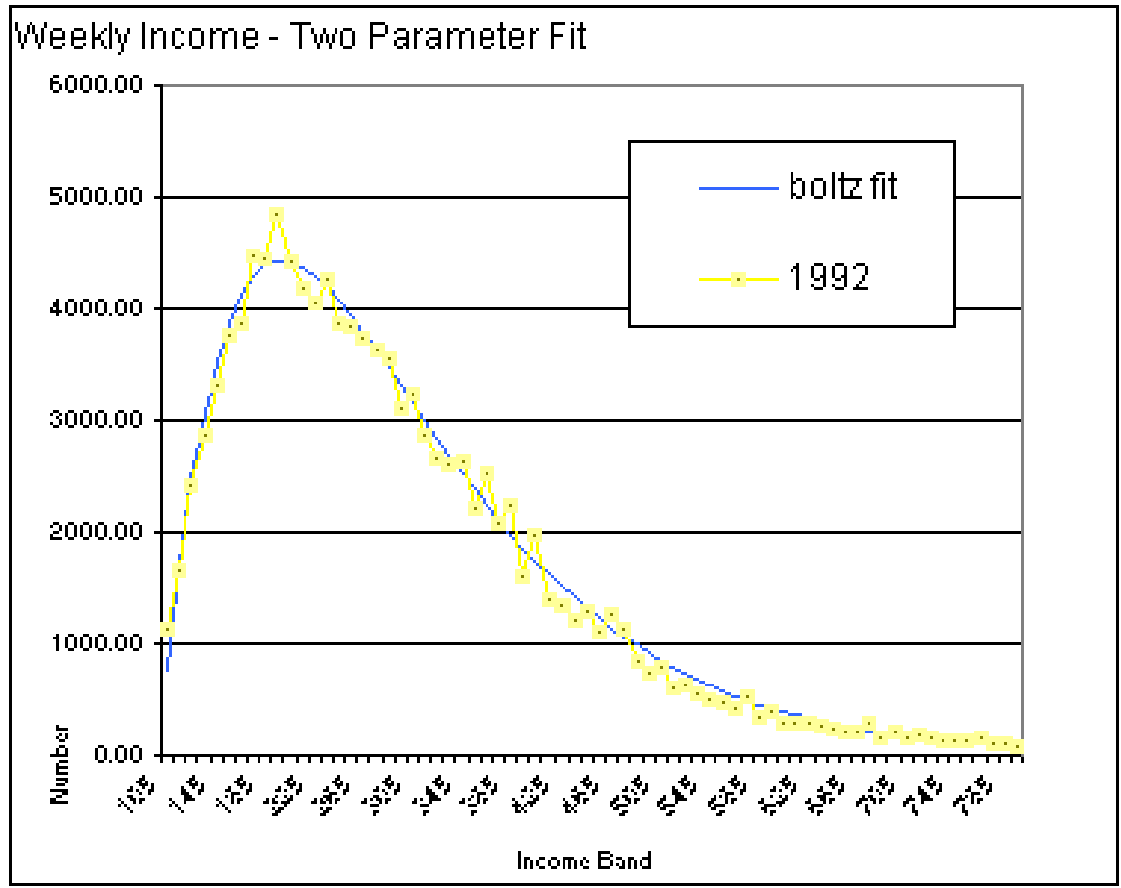

S

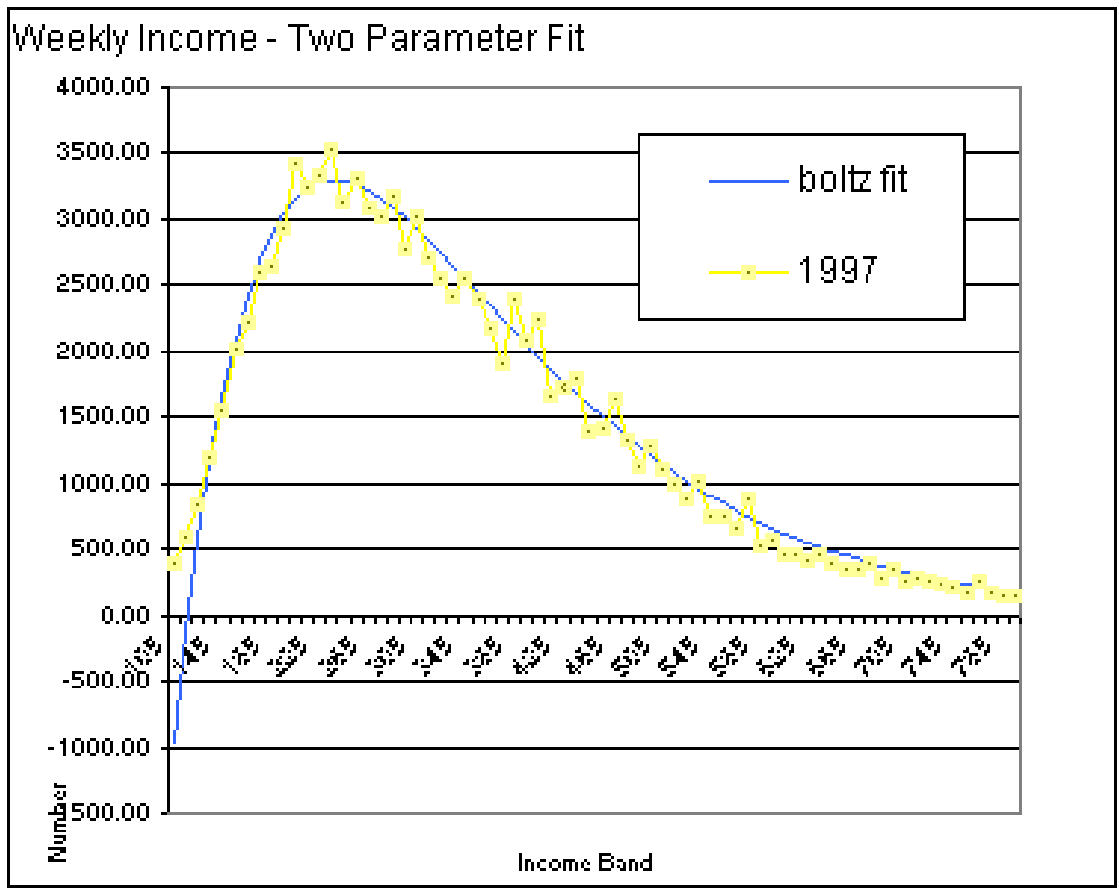

T

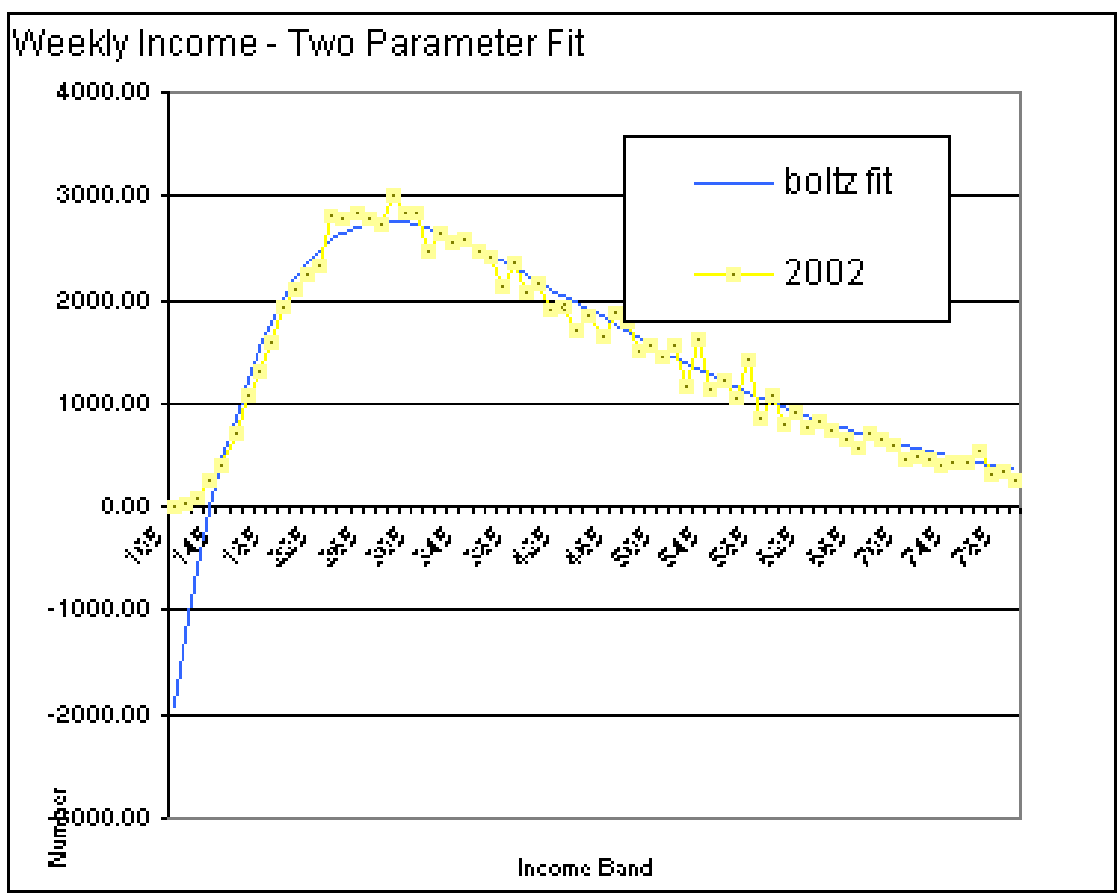

U

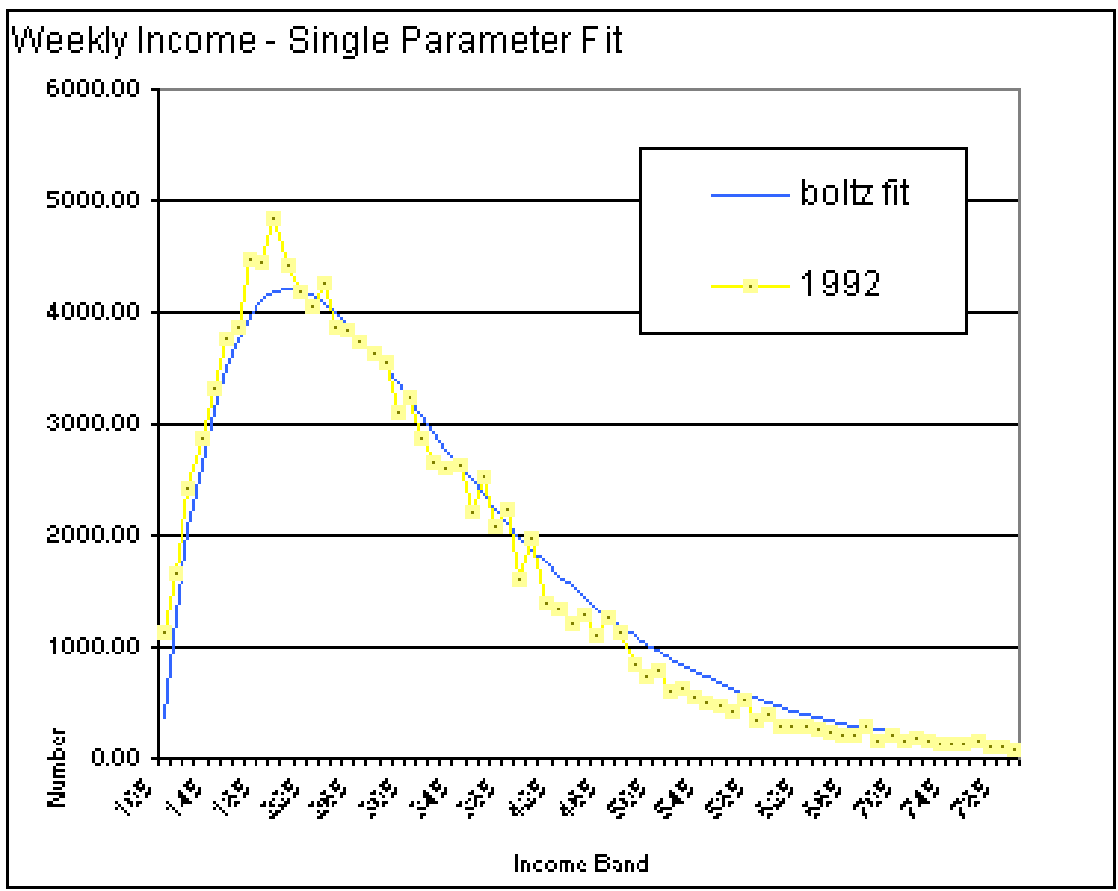

V

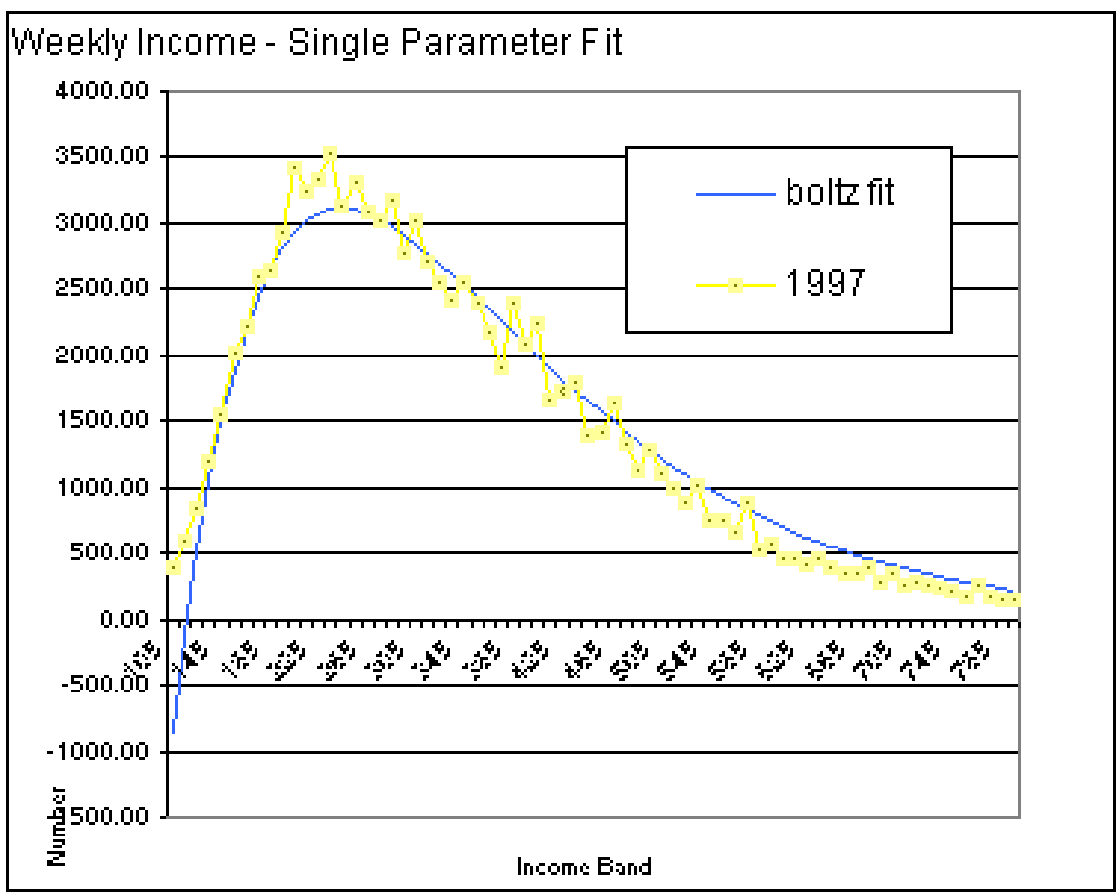

W

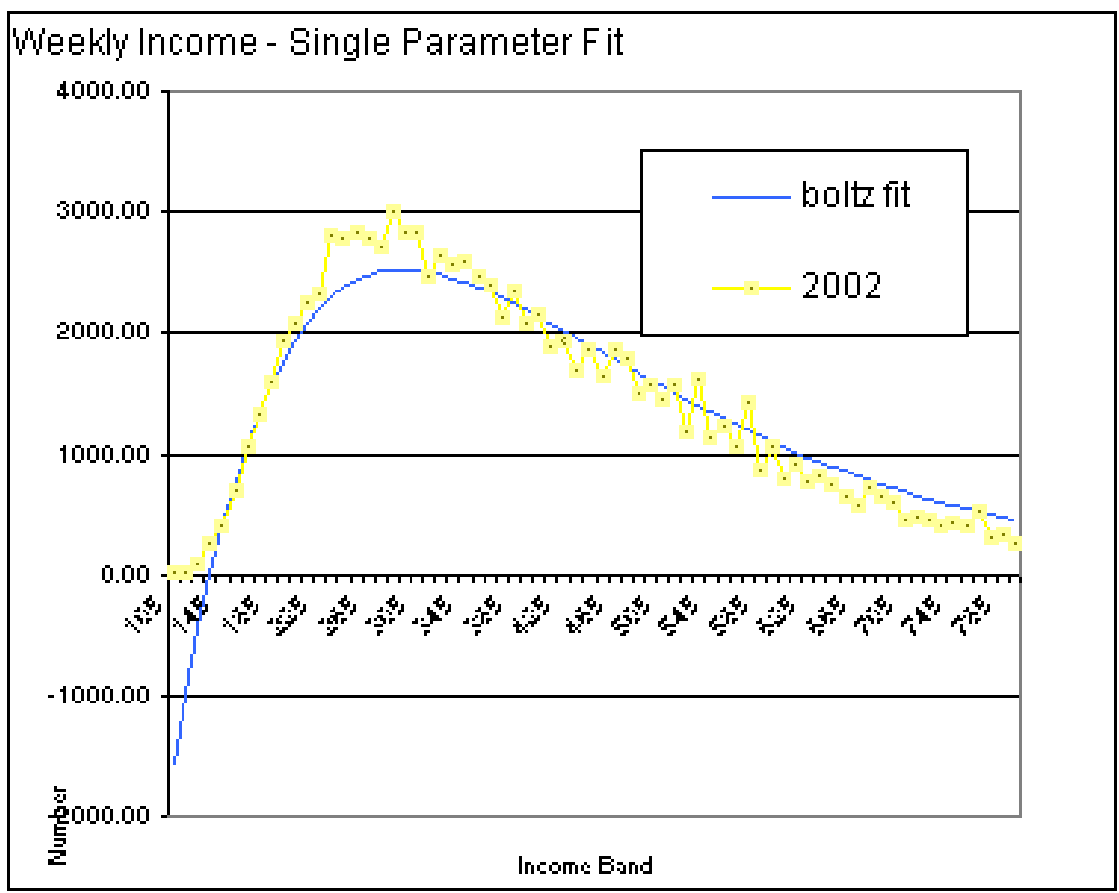

X

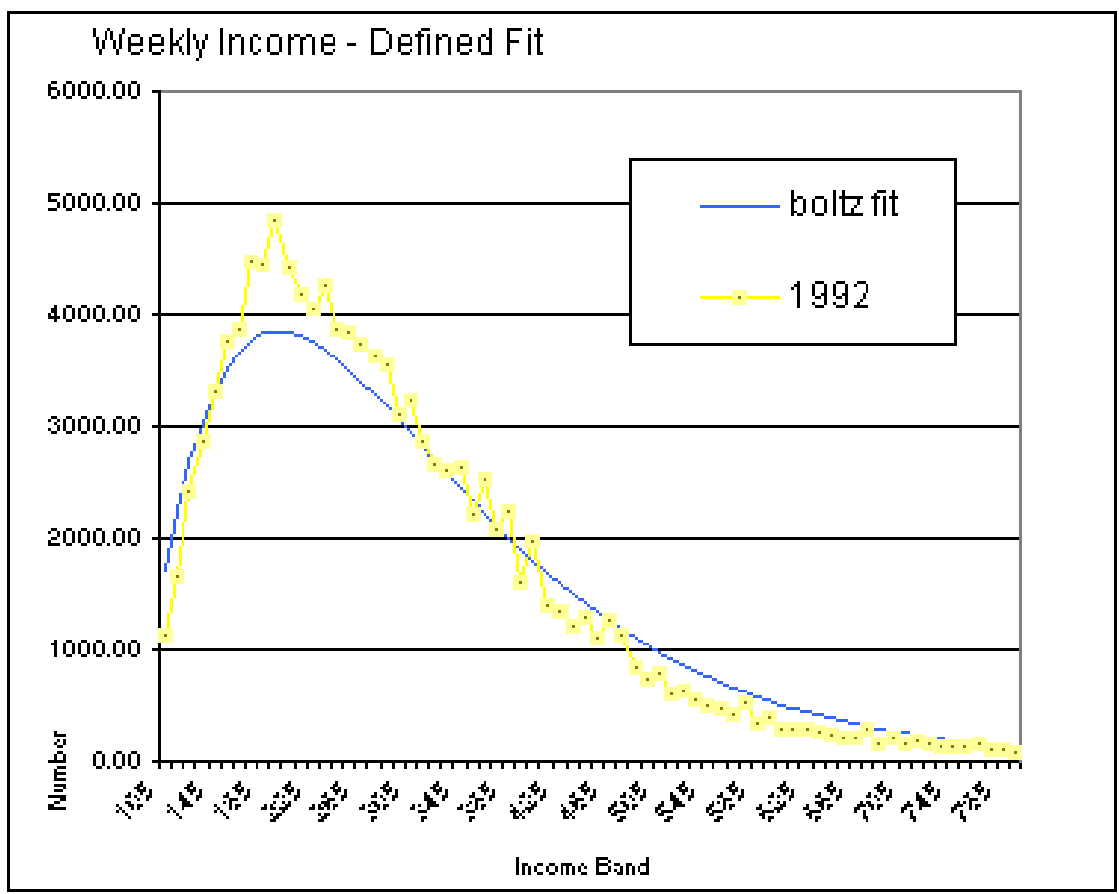

Y

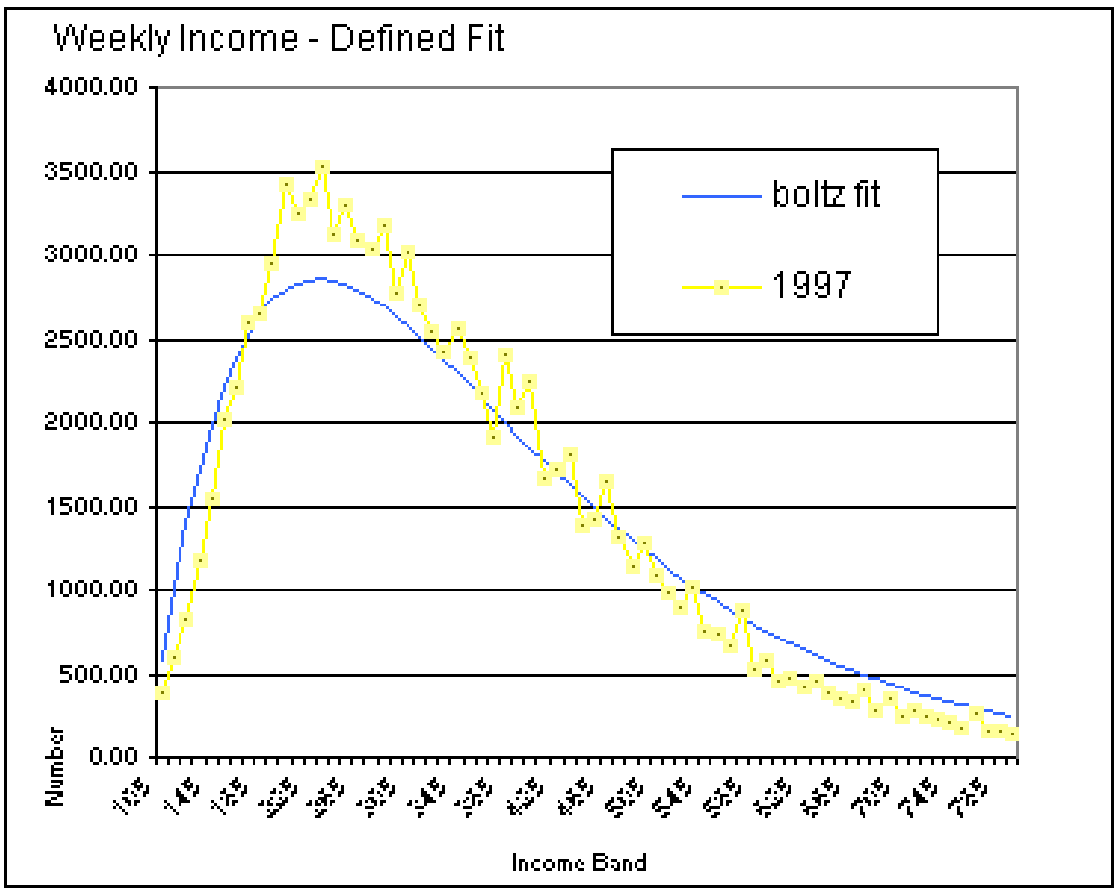

Z

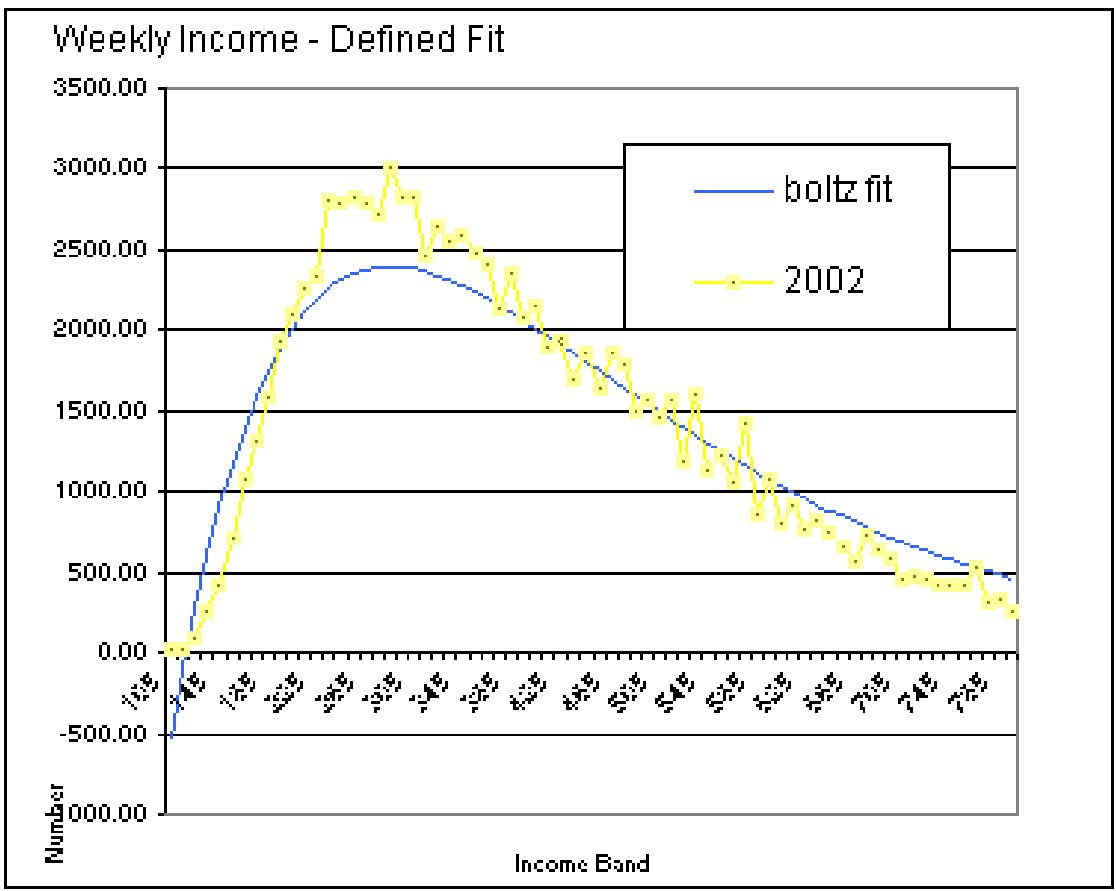

AA

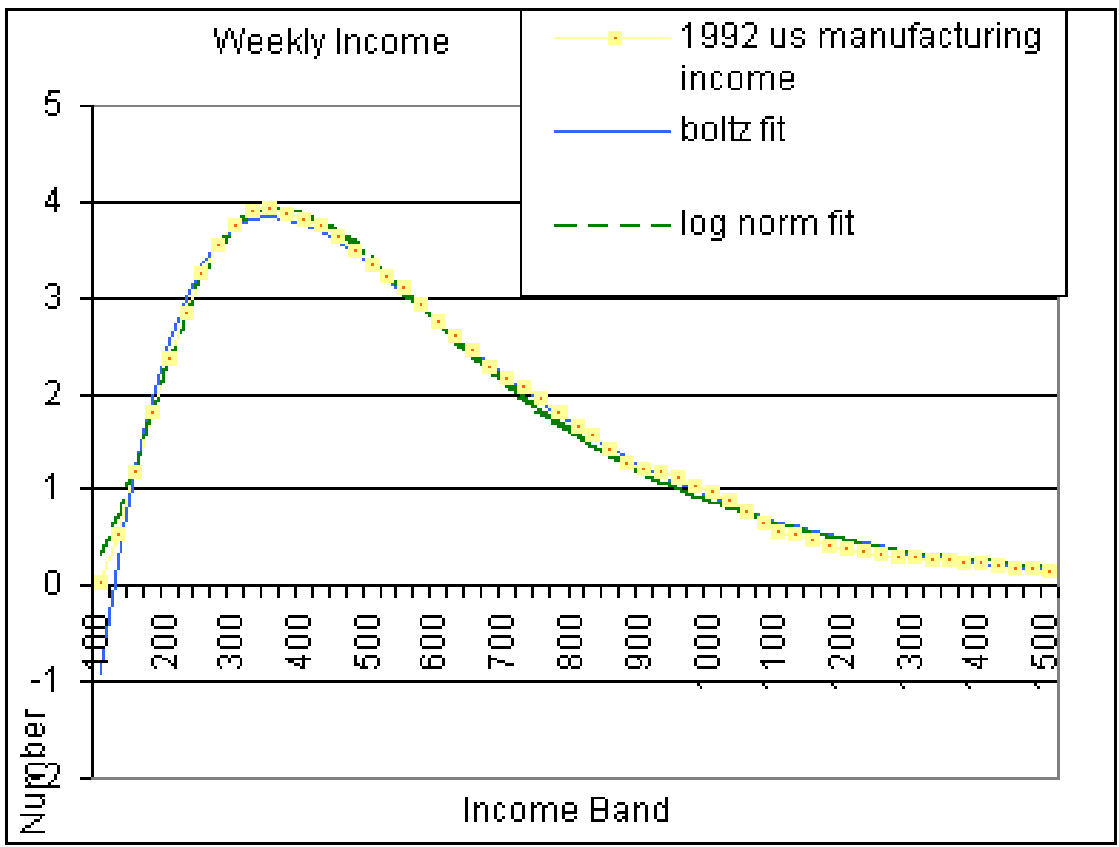

BB

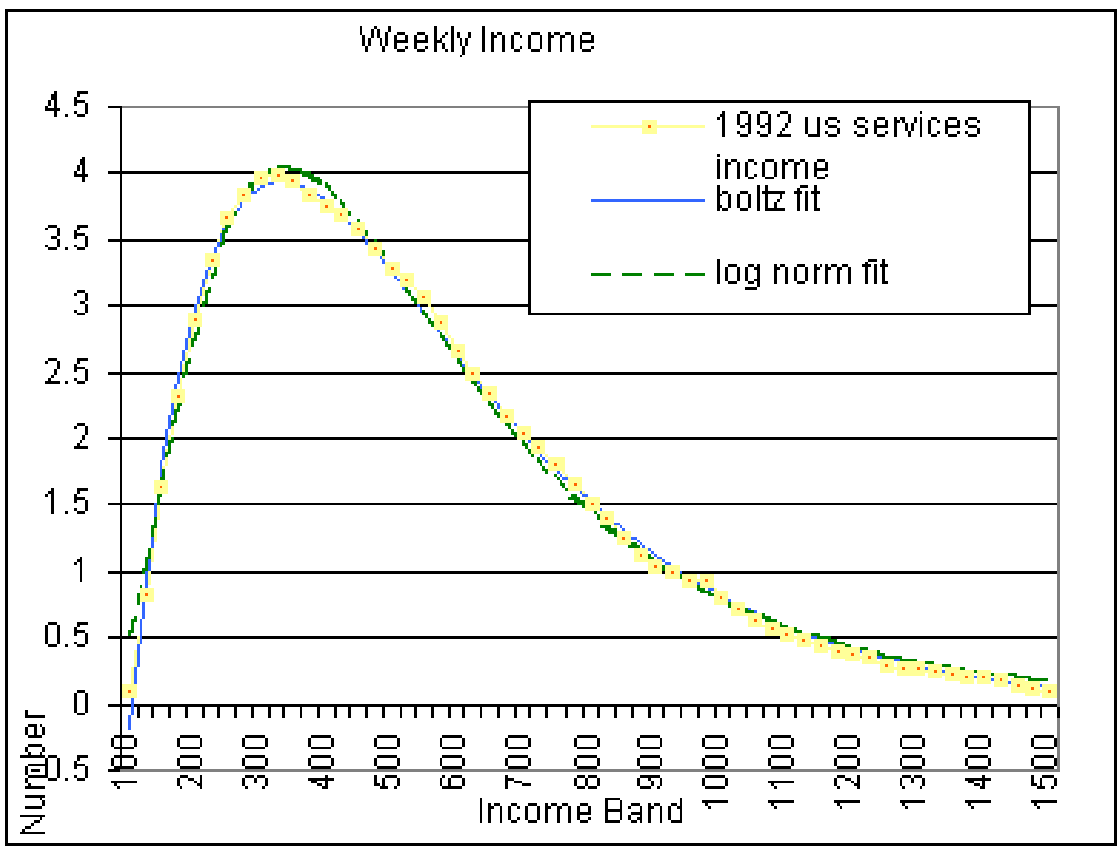

CC

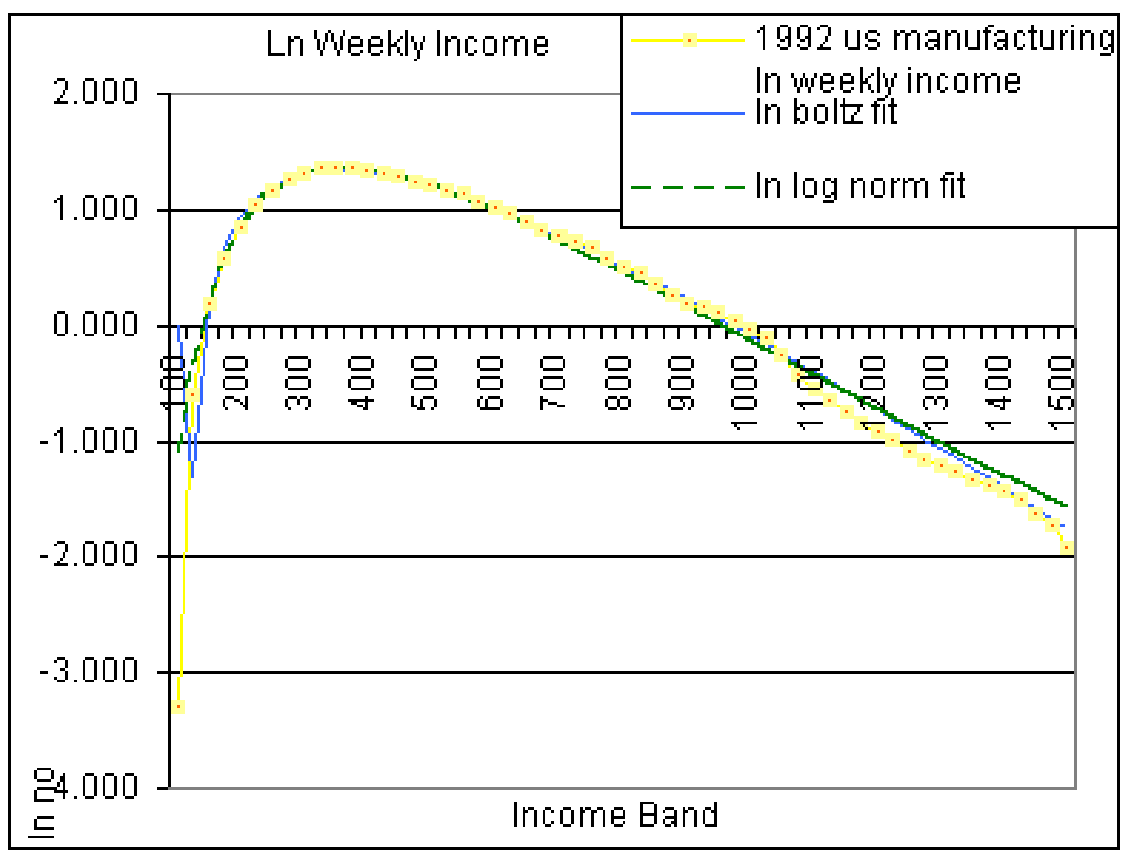

DD

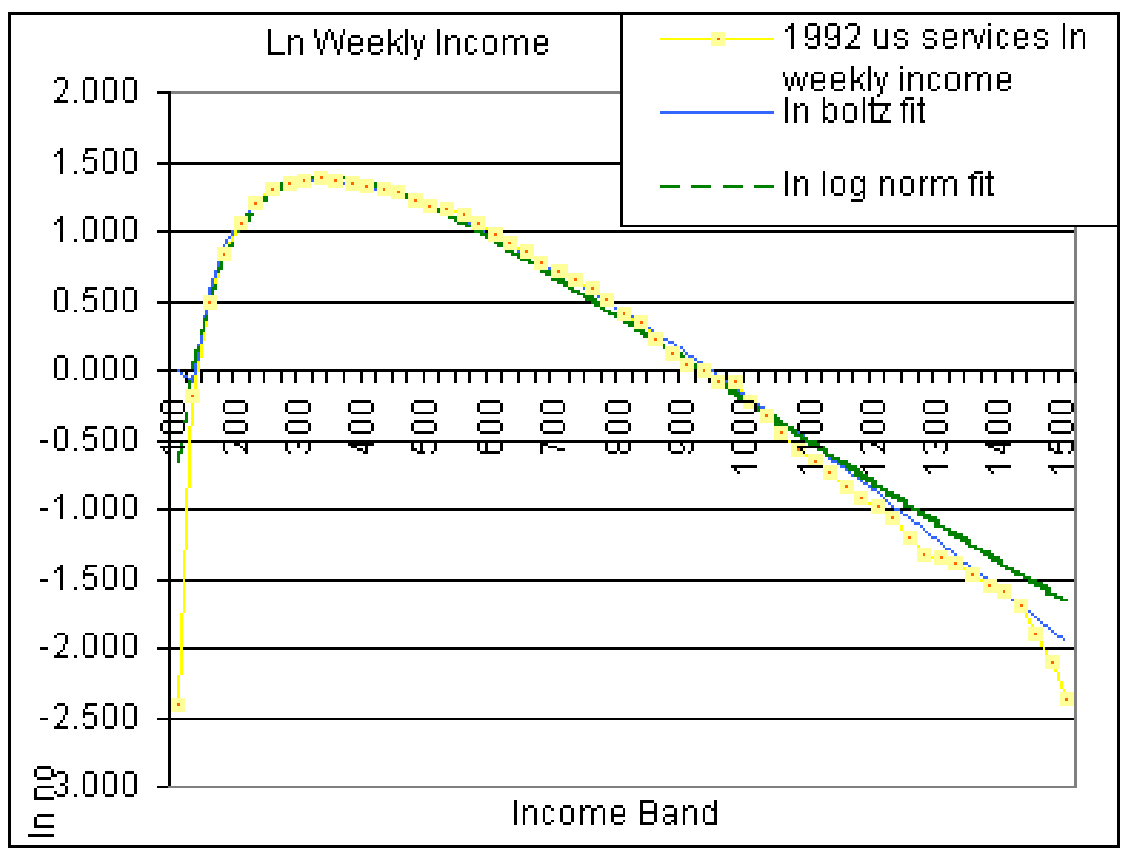

EE

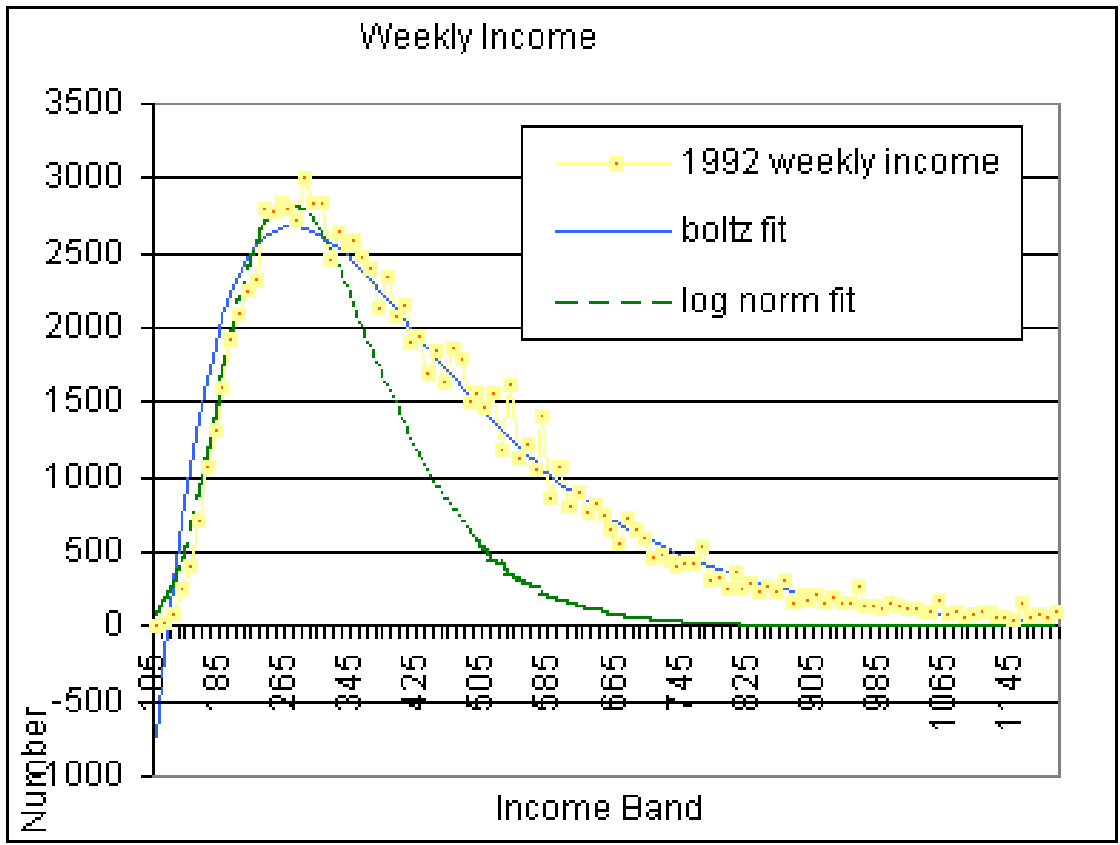